\documentstyle[prd,aps]{revtex}

\input epsf

\begin{document}

\draft
\title{Post-Newtonian SPH calculations of binary neutron star 
coalescence. II. Binary mass ratio, equation of state, and spin dependence.}
\author{Joshua A.\ Faber, Frederic A.\ Rasio, and Justin B.\ Manor}
\address{Department of Physics, Massachusetts Institute of Technology,
Cambridge, MA 02139}

\date{\today}

\maketitle

\begin{abstract}
Using our new Post-Newtonian SPH (smoothed particle hydrodynamics)
code, we study the final coalescence and merging of neutron star (NS) binaries.
We vary the stiffness of the equation of state (EOS) as well as the
initial binary mass ratio and stellar spins.  
Results are compared to those of Newtonian calculations, with and
without the inclusion of the gravitational radiation reaction.  We find a
much steeper decrease in the gravity wave peak strain and luminosity with
decreasing mass ratio than would be predicted by simple point-mass
formulae.  For NS with softer EOS (which we model as simple $\Gamma=2$
polytropes)  we find a stronger gravity wave emission, with a
different morphology than for stiffer EOS (modeled as $\Gamma=3$
polytropes as in our previous work). 
We also calculate the coalescence of NS binaries with an irrotational initial 
condition, and find that the
gravity wave signal is relatively suppressed compared to the
synchronized case, but shows a very significant second peak of
emission.  Mass shedding is also greatly reduced, and occurs via a
different mechanism than in the synchronized case.  We discuss the
implications of our results for gravity wave astronomy with laser
interferometers such as LIGO, and for theoretical models of gamma-ray
bursts (GRBs) based on NS mergers.
\end{abstract}


\section{Introduction and Motivation}\label{sec:intro}

Coalescing neutron star (NS) binaries are among the most important sources
of gravitational radiation for detection by LIGO \cite{1},
VIRGO \cite{2}, GEO \cite{3,3a}, and TAMA \cite{4}. 
These interferometers will be most
sensitive to gravity waves in the frequency range from $\sim 10\,{\rm
Hz}$ to $\sim 300\,{\rm Hz}$, corresponding to the last few thousands
of orbits prior to final merging. During final merging, the characteristic
frequency is $\gtrsim 1\,{\rm kHz}$, requiring the use of special
narrow-band detectors \cite{15}, which are currently
being tested on the GEO~600 interferometer in Germany \cite{3a}.  
The detection of these merger wave forms, combined with theoretical
knowledge about the hydrodynamics of the merger process, could yield detailed 
information about the behavior of the NS fluid, and, in particular, the EOS of 
nuclear matter at high density. This is the second of a series of
papers in which we attempt to develop this theoretical knowledge using
3D numerical hydrodynamics calculations of NS mergers in
post-Newtonian (PN) gravity. 

The first Newtonian calculations of binary NS coalescence were
performed more than 10 years ago by Nakamura and Oohara using a 
grid-based Eulerian finite-difference code \cite{Nak1,Nak2}.
Rasio and Shapiro (Ref.~\cite{RS13}, hereafter RS1--3 or collectively
RS) later used Lagrangian SPH (smoothed particle hydrodynamics)
calculations to study the stability properties of close binary NS
configurations in strict hydrostatic equilibrium, and follow the
evolution of unstable systems to complete coalescence.
RS demonstrated for the first time that, even in Newtonian gravity,
an innermost stable circular orbit (ISCO) is imposed by
global hydrodynamic instabilities, which can drive 
a close binary system to rapid coalescence once the tidal interaction 
between the two NS becomes sufficiently strong.
Since then many other groups have performed Newtonian calculations
of binary NS coalescence, concentrating on various aspects of the problem. 
Most recent calculations have used either SPH
(see, e.g., \cite{Zhu1,Zhu2,Dav,Ros1,Ros2}), or 
PPM (the Eulerian, grid-based, piecewise parabolic method;
see, e.g., \cite{New,Swe,RJS,RJTS,RRJ}).
Refinements have included the addition of terms to approximate energy
and angular momentum losses to gravitational radiation (see
\cite{Nak2,Ros2,RJS}) and more detailed treatment of the nuclear physics
(e.g. \cite{Ros1,RJTS}, with applications to models of gamma-ray bursts).

The first calculations to include the lowest-order post-Newtonian (1PN)
corrections to Newtonian gravity, as well as the lowest-order dissipative
effects of the gravitational radiation 
reaction (2.5PN) were performed by Oohara and  
Nakamura using an Eulerian grid-based method \cite{Nak3}.
More recently, Faber and Rasio (\cite{FR1}, hereafter Paper~1), 
and Ayal et al. \cite{Ayal},
have performed SPH calculations that included all 1PN and 2.5PN
corrections,
using a PN formalism developed by Blanchet, Damour, and Sch\"afer
(\cite{BDS}, hereafter BDS).

These calculations revealed that the addition of 1PN corrections to
the hydrodynamic equations can have a marked effect on the gravity
wave signal.  
In Paper~1, we presented a detailed comparison between two coalescence 
calculations (for initially synchronized binaries containing two identical
stars modeled as simple $\Gamma=3$ polytropes), one including both 1PN and
2.5PN corrections, and the other including only Newtonian gravity plus
radiation reaction (2.5PN corrections). 
We found that the 1PN effects produce several luminosity peaks in the
gravity wave 
emission, whereas the Newtonian merger shows a simpler signal with a single
peak followed by exponential damping of the emission.
In addition, 1PN effects accelerate the final coalescence and produce large 
tidal lag angles for both NS just prior to contact, leading to a more
``off-center'' 
collision.  This produces a complex pattern of quadrupole oscillation and
differential rotation in the merger remnant, accompanied by a characteristic 
modulation of the gravity wave signal.  
 
Several groups have been working on fully general relativistic (GR)
calculations of NS mergers, combining the techniques of 
numerical relativity and numerical hydrodynamics in 3D \cite{16,Shi1,Shi2}.  
However, this work is still in the early stages of development, 
and only preliminary results have been reported so far.  
Obtaining accurate gravitational radiation wave forms
from full GR simulations is particularly difficult, since the waves
must be extracted at the outer boundaries of large 3D grids extending
out into the true wave zone of the problem.
However, fully GR calculations are essential for addressing the
question of the 
stability to gravitational collapse and ultimate fate of a merger.
Using a GR formalism developed by Shibata \cite{Shi1}, Shibata and Uryu
\cite{Shi2} find that
collapse to a black hole on a dynamical time scale during final
coalescence does not happen for NS with stiff EOS and realistic
parameters. Only with unrealistically compact initial NS models
(starting already very close to the maximum stable mass) does 
collapse to a black hole occur on the merger time scale.
This confirms our expectations from the discussion in Paper~1:
for realistic NS with stiff EOS, PN calculations of binary coalescence
can provide results that remain at least qualitatively correct  all the
way to a complete merger.

In this paper, we continue the work begun in Paper~1, using our PN 
SPH code to study the coalescence of NS binaries.
Here, we relax some of the simplifying assumptions made in Paper~1
in constructing the initial models for NS binary systems. We no longer
assume the binary system to contain two identical stars, and perform 
several calculations for binary mass ratios $q\ne1$. We also allow for
a nonsynchronized initial configuration, instead of the rigidly rotating
binary system assumed in Paper~1. Finally, we consider NS with
somewhat softer EOS, but still modeled as simple polytropes.
In the rest of this section, we discuss the theoretical and observational
constraints on these various assumptions.

Mass measurements for NS derived from the timing
of relativistic binary radio pulsars are all consistent with a
remarkably narrow underlying Gaussian distribution with
$M_{NS}=1.35\pm0.04\,M_\odot$ \cite{Tho}.
The largest observed departure from $q=1$
in any known binary pulsar with
likely NS companion is currently $q=1.386/1.442=0.96$
for the Hulse-Taylor pulsar PSR B1913+16 \cite{Tay}.
For comparison, $q=1.349/1.363=0.99$ for PSR B2127+11C 
\cite{21},
and $q=1.339/1.339=1$ for PSR B1534+12 \cite{Tho,Wol}.
 Although the equal-mass case is
clearly important, one should not conclude from these observations
that it is unnecessary to consider coalescing NS binaries with
unequal-mass components. Indeed, it cannot be excluded that other 
binary NS systems (that may not be observable as binary pulsars) 
could contain stars with significantly different masses, especially
considering the different histories of the primary and secondary in
most binary NS evolution scenarios. Theoretically, the predicted NS
masses from stellar evolution and supernova calculations span a
considerably wider range than observed \cite{NSmass}.
Moreover, Newtonian calculations of binary NS coalescence (RS2, \cite{Zhu2}) 
have shown
that even very small departures from $q=1$ can drastically affect
the hydrodynamics. A small asymmetry in the initial system grows
rapidly during the merger, and there is almost no disruption of the 
primary (more massive star), which simply sinks to the center 
of the merger remnant (RS2).  
A more extreme asymmetry, such as $q\simeq 0.5$, can lead to
a brief episode of mass transfer from the secondary (less massive star) 
onto the primary, followed by a sudden widening of the
binary system, rather than a merger (RS2).  The same behavior is seen
in hydrodynamic simulations of coalescing binaries containing a NS in orbit
around a more massive black hole  (see \cite{JERF,LK}).
In this paper, we will not consider such extreme mass ratios, but will
focus instead on covering well a narrower but realistic range of values 
from $q=0.8$ to $q=1$ (For comparison, previous, purely Newtonian
calculations have been performed for $q=0.5$ and $q=0.85$; see RS2, \cite{Zhu2}).

Because of the low viscosity of the NS fluid,
the tidal synchronization time for NS binaries is always considerably
longer than the orbital decay time scale. Therefore, it is expected that
the NS should be spinning slowly prior to final coalescence \cite{Irrot}.
In the limit of nonspinning stars at large separation, the binary
configuration just prior to final coalescence will be {\it irrotational\/},
i.e., the fluid motion inside each star will have zero vorticity, although 
it can have a small amount of angular momentum (\cite{Irrot,LRS1,LRS3}).
In contrast, in many previous studies of NS binary coalescence (and in
Paper~1), 
{\it synchronized\/} initial conditions have been used for simplicity 
(RS2, \cite{Nak1,Nak2,New,Swe,Nak3}). For a synchronized binary, the entire
mass of fluid is in rigid rotation, and an equilibrium initial configuration
can be constructed easily using relaxation techniques in the corotating frame
of the binary (where the problem reduces to hydrostatics). 
These relaxed initial conditions, while somewhat unphysical, are much less 
susceptible to spurious oscillations of the fluid around the correct
quasi-equilibrium configuration (RS).
In this paper, we will present a method for constructing accurate 
irrotational initial conditions, and we will compare the results of PN
calculations 
for the coalescence of irrotational NS binaries to those obtained in Paper~1 
for initially synchronized NS binaries.

Here, as in Paper~1, we model NS with a stiff EOS as simple 
polytropes, i.e., the pressure $P$ is a function of the density only, of
the form $P=kr_*^\Gamma$, where $k$ is called the polytropic constant,
$\Gamma$ is the adiabatic exponent, and
$r_*$ is the rest mass density of the fluid (we adopt the notation
of BDS), not to be confused with the total mass-energy density 
of the fluid \cite{Note1}.  

A polytropic EOS provides a convenient approximation to many different
realistic NS EOS calculated under various assumptions about the
still uncertain microphysics above nuclear density. Higher values of
$\Gamma$ represent stiffer EOS, while lower values represent softer EOS.
In addition, it allows for direct comparison to previous Newtonian 
work, which was done mainly for polytropes.
The most commonly adopted polytropic EOS for NS uses $\Gamma=2$
(RS1, \cite{Zhu1,New,Swe}). 
Newtonian polytropes with $\Gamma=2$ have a radius that is independent of mass,
a well-known property satisfied approximately by all realistic NS models
(\cite{ST}).  Several tabulated NS
EOS are found to have effective adiabatic exponents that are close to
this value, or slightly higher. In particular, the Lattimer-Swesty
EOS \cite{Lat}, used in several previous calculations
\cite{Ros1,Ros2,RJS,RJTS,RRJ} has an effective
adiabatic index $\Gamma\simeq2.3$ for a $1.4\,M_\odot$ NS, assuming
the lowest available value of the nuclear compressibility
($K=180\,$MeV).  The latest microscopic NS EOS, constrained by nucleon
scattering data and the binding of light nuclei, and incorporating
three-body forces, are even stiffer, with $\Gamma\simeq 3$ (see, e.g., 
\cite{Bay} for a summary, \cite{Akm} for a recent update, and
\cite{LRS3}  for polytropic fits). In this paper, we will present and compare
calculations performed for $\Gamma=3$ (as in Paper~1), and for $\Gamma=2$.
We feel that this range corresponds reasonably to the theoretical uncertainty
in recent determinations of the NS EOS.
We will {\it not\/} consider in this paper the much softer EOS that may result 
from even more uncertain considerations of exotic states of matter
such as pion condensates or strange quark matter \cite{Kaon}.  
Note that PN expansions are more
difficult to do consistently with softer EOS, since the stars are more
centrally condensed and thus the central gravitational field 
is much stronger. There is some advantage to using
softer EOS within the framework of the BDS formalism, however, since
smaller values of $\Gamma$ help keep several 1PN quantities small 
(see Paper~1, Eqs.~A10-12).
A key difference between stiffer and softer EOS is captured by our range
of adiabatic exponents. Indeed, previous Newtonian work has shown a significant
difference between $\Gamma=2$ and $\Gamma=3$,
since polytropes with $\Gamma\gtrsim 2.4$ can support a
stable, rapidly rotating {\it triaxial\/} configuration, which would lead to a
persistent gravity wave signal, whereas those with $\Gamma\lesssim
2.4$ generally damp out to oblate {\it spheroids\/}, which have no
time-dependent 
quadrupole moment and therefore cannot radiate gravity waves (RS2).
The existence 
of a long-lasting exponentially damped gravity-wave emission tail would have a
significant effect on the integrated power spectrum of the
gravitational radiation signal from binary coalescence.

The outline of our paper is as follows. 
Section~II presents our numerical methods, including a brief description
of our PN SPH code, and the procedure used to construct
irrotational initial conditions (The construction of synchronized
initial conditions and the SPH method are discussed in more details
in Paper~1).  Additionally, we describe the parameters and assumptions
for all runs quoted in this paper. 
Section~III presents our numerical results, focusing in turn on the effects of
varying the EOS, the mass ratio, and the initial spins, as well as 
the differences between Newtonian runs and PN runs.
Motivation for future work as well as a brief summary of our
results to date are presented in Section~IV.
 
\section{Numerical Methods and Summary of Calculations}

\subsection{The PN SPH code}

All our calculations were performed using the post-Newtonian (PN) smoothed
particle hydrodynamics (SPH) code described in detail in Paper~1.  It is a
Lagrangian, particle-based code, which has been adapted to the PN
formalism of BDS.  As in Paper~1, we use a hybrid 1PN/2.5PN formalism, in
which 1PN corrections are artificially scaled down, so as to remain
numerically tractable, while 2.5PN effects are included for realistic NS 
parameters. Throughout this paper we use units such that $G=M=R=1$,
where $M$ and 
$R$ are the rest mass and radius of the primary NS. 
In these units, there are then two different values of the speed of light $c$
corresponding to the 1PN and 2.5PN corrections.  As in
Paper~1, we take $c_{2.5PN}=2.5$,
which gives $GM/Rc_{2.5PN}^2=0.16$, corresponding to realistic NS parameters,
but we take $c_{1PN}=4.47$, giving $GM/Rc_{1PN}^2=0.05$, to keep all 1PN 
quantities sufficiently small.

Models for the primary NS with a $\Gamma=3$ polytropic EOS
(i.e., $P=kr_*^3$) were taken from Paper~1. To construct models for the
secondary NS when $q \ne 1$, we kept the same value of the polytropic
constant $k$
determined for a primary with $GM/Rc^2=0.05$, but we computed models
with a lower mass using a relaxation technique.  
Note that, in contrast to their Newtonian counterparts, PN polytropes
do not obey a simple power-law mass-radius relation, and therefore models
for the secondary cannot be obtained simply by rescaling models for
the primary. 
Additionally, we computed a series of
$\Gamma=2$ polytropic models for single NS with values of $GM/Rc^2$ between
$0.01$ and $0.05$, as well as lower-mass models based on the primary model
with $GM/Rc^2=0.05$. While Newtonian polytropes with
$\Gamma=2$ have a radius independent of the
mass (for fixed $k$), the radius of a PN $\Gamma=2$ polytrope increases with 
decreasing mass. 
The properties of the $\Gamma=2$ PN polytropic sequence are summarized in 
Table~\ref{table:single} and the properties of NS models with lower masses
are listed in Table~\ref{table:single2}.

To set up synchronized initial conditions, we used the same relaxation scheme 
described in Paper~1 to create accurate equilibrium configurations.  
The two NS are initially placed along the $x$ axis, with a binary 
separation $r_0$ (distance between the centers of mass of the two stars), 
and the binary center of mass at the origin. The orbital
motion is assumed to be in the $x-y$ plane, and the orbital angular momentum
vector in the positive $z$ direction. We assume that all SPH particle
velocities ${\bf v}_i$ vanish in the corotating frame of the binary, 
and we use an iterative
method to find the proper force on each particle in equilibrium, holding 
the binary separation fixed.  As discussed in Paper~1, there are subtle 
differences between the particle velocities ${\bf v}_i$ and momenta
${\bf w}_i$, 
which must be treated with care (see Paper~1, Eqs.~A23--24, and Eq.~B8).
The angular velocity $\Omega$ of the corotating system is
calculated from the centripetal accelerations of the two components
of the binary, 
\begin{equation}
\Omega=\sqrt{\frac{-\dot{v}^{(1)}_{x}+\dot{v}^{(2)}_{x}}{2r_0}}.
\label{eq:omega}
\end{equation}
Here ${\bf v}^{(1)}$ and ${\bf v}^{(2)}$ are the center-of-mass velocities of the
primary (located on the positive $x$ side) and secondary (located on the
negative $x$ side), respectively.
As in Paper~1, we set the initial velocity field of the fluid to be
$v_x=-\Omega y$, $v_y=\Omega x$ when the dynamical run is started
(in the inertial frame). 

\subsection{Irrotational initial conditions}\label{sec:irrot}

Since there is no simple way to relax an irrotational
binary configuration, to strict equilibrium,
we used the results of Lombardi, Rasio, and
Shapiro (\cite{LRS}, hereafter LRS)
 to construct approximate initial conditions.  LRS calculated
PN equilibrium solutions for irrotational binary NS with polytropic EOS, assumed to
have self-similar ellipsoidal density profiles, with the density as a
function of radius given by the 1PN expansion of the 
Lane-Emden equation.  Approximate
solutions were determined by minimizing the total energy of the binary
configuration, including 1PN terms. The resulting NS models are the compressible,
PN analogues of the classical Darwin-Riemann ellipsoids for incompressible
fluids \cite{LRS1,Chandra,Note2}. 

In this paper we will only consider a single irrotational binary model,
with $q=1$ and $\Gamma=3$.
To construct the initial binary model, we adopt the equilibrium shape 
of the stars determined by LRS, but in combination with
the internal structure of our single star models (calculated previously for
synchronized initial conditions).  Since our PN runs were started from a rather
large initial separation, $r_0=4.0\,R$, where the tidal deformations are
still extremely small, we find that the initial oscillations that result
from small departures from true equilibrium are negligible.  
For an EOS with $\Gamma=3$ ($n=0.5$, where $n=1/(\Gamma-1)$ is the
polytropic index),
an initial separation of $4.0\,R$, and a compactness $GM/Rc^2=0.05$, we
find from Table~III of LRS that the axes of the ellipsoidal figure for each
star are given by
$a_1/R\simeq 1.02$, $a_2/a_1 \simeq
0.96$, and $a_3/a_1\simeq 0.96$. Here $a_1$ is along the binary axis
($x$-direction), $a_2$ is 
in the direction of the orbital motion ($y$-direction), and
$a_3$ is along the rotation axis ($z$-direction).

The initial velocities of all SPH particles are calculated from
Eqs.~18-20 of LRS, which give
\begin{eqnarray}
v_x &=& -\Omega y(1-\frac{2a_1^2}{a_1^2+a_2^2})\\
v_y &=& \Omega x(1-\frac{2a_2^2}{a_1^2+a_2^2}).
\label{eq:virr}
\end{eqnarray}
It is easy to verify that this initial velocity field has zero 
vorticity in the inertial frame. We determine the angular velocity
from Eq.~\ref{eq:omega}, using an iterative method, since velocity
derivatives have a weak velocity dependence which makes it difficult
to find a self-consistent solution directly (see Paper~1, Eqs.~A17-A19
and Eq.~A23).  We find that three
iterations yields convergence to within a fraction of a percent, and
six to within one part in $10^4$.

\subsection{Summary of calculations}\label{sec:sumcalc}

We performed several large-scale SPH calculations of NS
binary coalescence for the equal-mass case, and a series of shorter runs
designed to study the effect of varying the mass ratio $q$.  
Table~\ref{table:runs} summarizes the relevant parameters of all runs
performed. Following the notation of Paper~1, we refer to all runs 
that included both 1PN and 2.5PN effects as ``PN
runs'', while those runs performed without 1PN corrections are
referred to as ``N runs.''
Note that {\it both\/} N and PN runs included the 2.5PN
gravitational radiation reaction effects. We did not perform any
new completely Newtonian calculations (except for a brief test run mentioned 
in Sec.~\ref{sec:eos}).

All runs use $5\times10^4$ SPH particles per NS (i.e., the total number of particles
$N=10^5$), independent of $q$. The number of SPH particle 
neighbors $N_N=100$.  Shock heating, which is normally treated
via the SPH artificial viscosity, was ignored, since it plays a
negligible role in binary coalescence, especially for fluids with a 
very stiff EOS (see Paper~1).  All Poisson
equations were solved on grids of size $256^3$, including the space
for zero-padding, which yields the proper boundary conditions.  All PN
runs were done using $1/c^2_{1PN}=0.05$ and $c_{2.5PN}=2.5$, as in Paper~1.  
Since we have an ambiguity in defining the relative time coordinate for
different runs, we redefine the initial time $t_0$ of each run in such a way 
that the time of the first gravity wave luminosity peak $t^{(1)}_{\rm max}=20$, 
as in run B1 (which starts from $t=0$). As a result, some runs started before 
$t=0$ (the starting time $t_0$ for each run is given
in the third column of Table~\ref{table:runs}).

Run A1 is identical to the N run described in Paper~1, for a binary
containing two identical NS modeled as $\Gamma=3$
polytropes. This run uses Newtonian gravity with radiation reaction 
(2.5PN) corrections, 
and started from an initial separation $r_0=3.1\,R$.
Runs A2, A3, A4, and A5
feature the same initial conditions, but have mass ratios $q=0.95$,
$0.90$, $0.85$, and $0.80$, respectively.  Unlike A1, in which the simulation
ran until a stable triaxial remnant was formed, runs A2--A5 were 
terminated after the completion of the first peak in the gravity wave
luminosity.

Run B1 is the PN run of Paper~1, also for two identical $\Gamma=3$ 
polytropes, but with 1PN and 2.5PN corrections included,
and an initial separation of $r_0=4.0R$.
Runs B2 and B3, like their Newtonian counterparts, correspond to
mass ratios $q=0.9$ and 0.8, respectively, and the same shorter 
integration time.

Run C1 is a Newtonian run (with radiation reaction effects) for two
identical $\Gamma=2$ polytropes.  Since the ISCO for this softer EOS is
located at smaller binary separation than for $\Gamma=3$, (see,
e.g., \cite{LRS}) we start this run with $r_0=2.9\,R$.  As a test, 
we compared the results of this run to one with all identical parameters
but with a slightly larger initial separation, $r_0=3.1\,R$, and found no 
measurable differences.  Run~C2 has a mass ratio of $q=0.8$ and the same EOS and
initial separation.  Runs~D1 and~D2 are the 1PN counterparts of C1 and C2,
with $\Gamma=2$, an initial separation $r_0=4.0\,R$,
and $q=1$ and $q=0.8$, respectively.  Run~C2 was terminated after
the first gravity wave luminosity peak, whereas run~D2 was continued until a
stable remnant configuration was reached, to allow for comparison with D1.

All the previous runs started from a {\it synchronized\/} initial binary
configuration, as in Paper~1. In contrast,
run~E1 started from an {\it irrotational\/} initial condition, with the
EOS, mass ratio, and initial separation as in run B1, and including all
1PN and 2.5PN terms.  
It was continued until a stable remnant configuration was reached.

\section{Results}\label{sec:results}

\begin{figure}
\centering \leavevmode \epsfxsize=6in \epsfbox{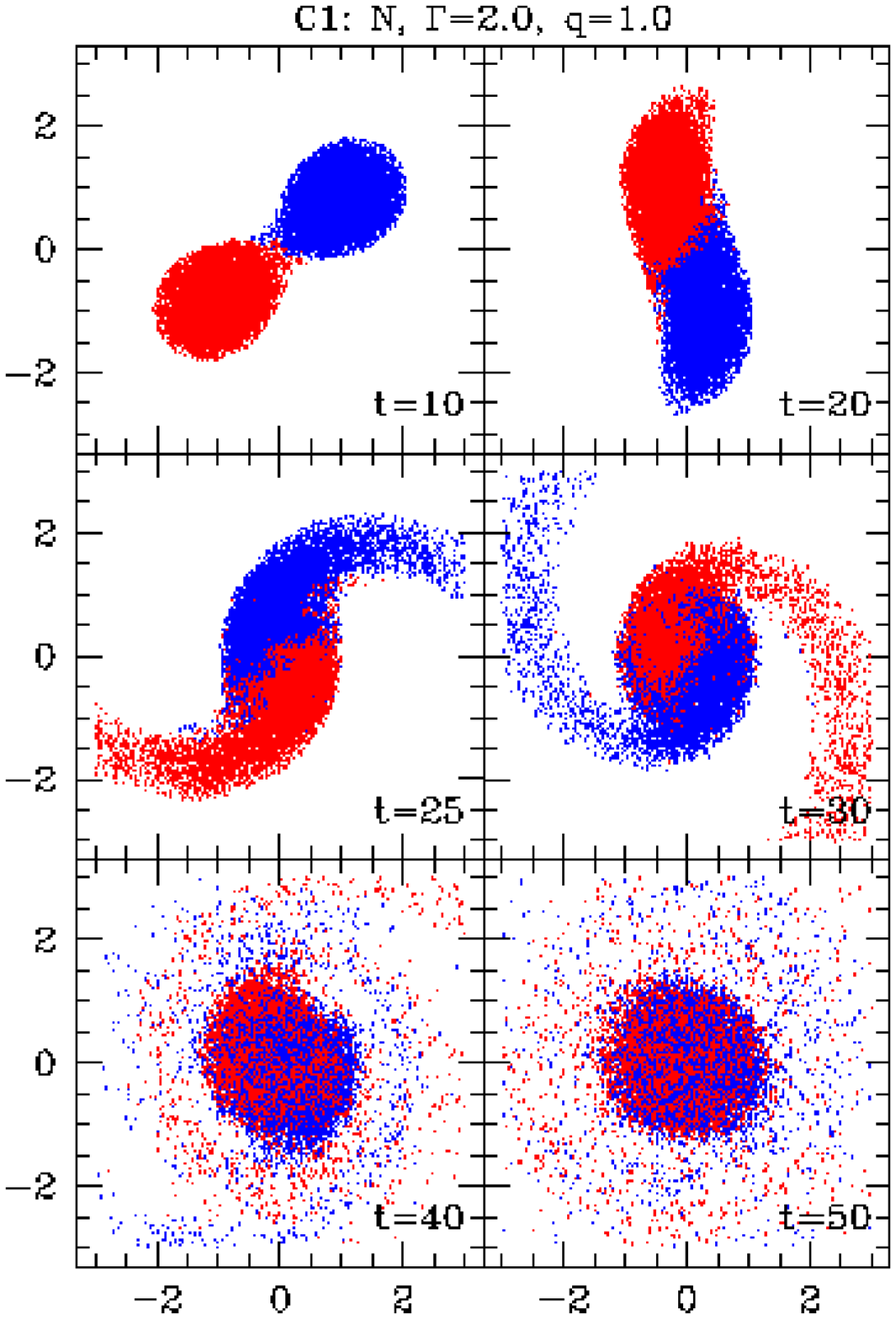}
\caption{Evolution of the system for Run C1 with $\Gamma=2$ and
$q=1$.  This run includes Newtonian gravity and radiation reaction
(2.5PN) corrections.  Projections of a
random subset of $20\%$ of all SPH particles onto the orbital (x-y)
plane are shown at various times.  The orbital motion is
counter-clockwise.  Units are such that $G=M=R=1$, where $M$ and $R$
are the mass and radius of a single, spherical NS.  Time has been shifted
so that the maximum gravity wave luminosity occurs at $t=20$.}
\label{fig:xy2n}
\end{figure}

\begin{figure}
\centering \leavevmode \epsfxsize=6in \epsfbox{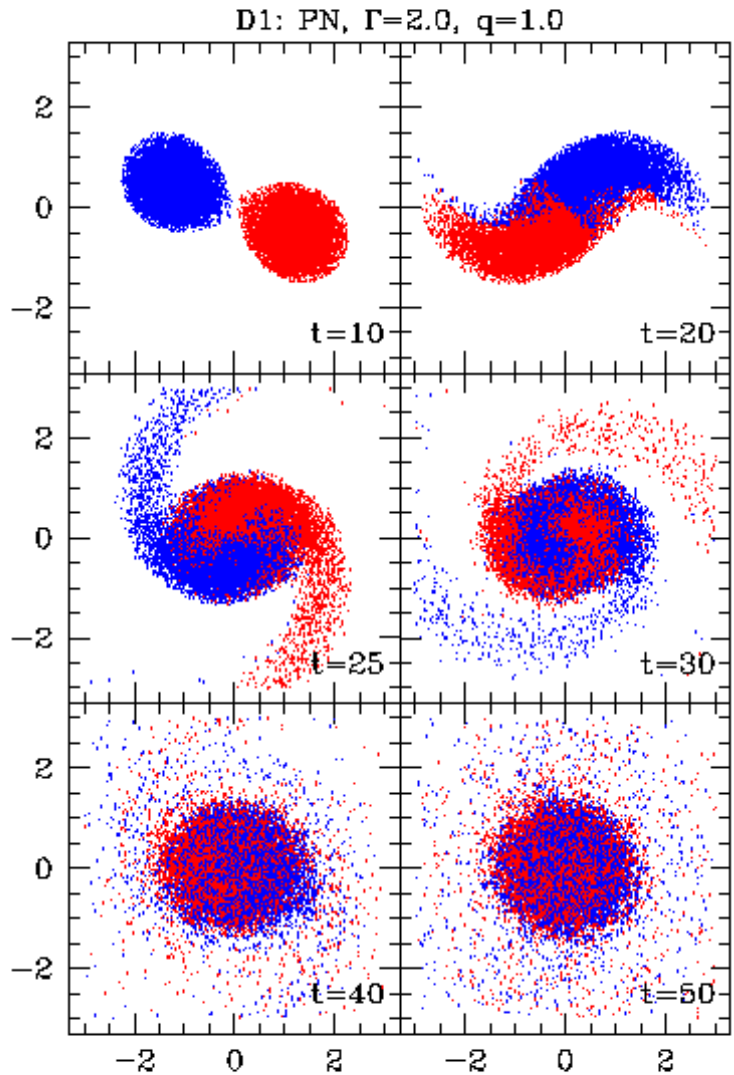}
\caption{Evolution of the system for Run D1, with $\Gamma=2$ and
$q=1$.  Here, all 1PN and 2.5PN corrections are included in the
calculation.  Conventions are as in Fig.~\protect\ref{fig:xy2n}.}
\label{fig:xy2p}
\end{figure}

As in Paper~1, we begin with a brief qualitative discussion of the
coalescence process, but focus here on the softer EOS with $\Gamma=2$.
In Figs.~\ref{fig:xy2n} and~\ref{fig:xy2p}, we compare
the evolution of the system in runs C1 and D1,
respectively.  Both runs are for two identical NS 
with a $\Gamma=2$ EOS, 
but run D1 includes 1PN effects whereas run C1 does not.  These plots
can be directly 
compared to Figs.~1 and~2 of Paper~1, which show the evolution of
runs A1 and B1, respectively (for two identical NS with $\Gamma=3$).
As seen in Fig.~\ref{fig:xy2p}, a significant 
tidal lag angle $\theta_{lag}$ develops in the PN system just prior to
final merging.  Values of $\theta_{lag}$ (calculated as the angle between the 
binary axis and the principal axis of each component) upon first contact 
are given in the fourth column of
Table~\ref{table:runs} for all runs.  Note that for binaries with
mass ratios $q\ne 1$, the secondary develops a larger lag angle than
the primary.  This effect is much more pronounced with the addition of 1PN
corrections.  In all cases, the merger is accompanied by mass shedding and 
the formation of spiral
arms, which may have a significant width.  These spiral arms dissipate
quickly (by $t\simeq 40$ in Figs.~\ref{fig:xy2n} and~\ref{fig:xy2p})
and merge to form a halo of material around the
axisymmetric, oblate remnant at the center of the system.
It should be noted that when we speak of ``mass shedding''
in this work (both Paper~1 and here), we mean that matter is ejected from the
central dense core into an outer halo, but {\it not to infinity\/}.
By the end of our simulations, all the matter forming the outer halo is 
still gravitationally bound to the system.

\subsection{Dependence on the NS EOS}\label{sec:eos}

As already pointed out in Paper~1, our PN calculations of NS binary 
coalescence are most
relevant for stiff NS EOS, for which most recent calculations give
values of $GM/Rc^2\simeq 0.1-0.2$ (for $M\simeq 1.5\,M_\odot$).  
Even if the true NS EOS were much softer, making strong GR effects dominant 
throughout the final binary coalescence, performing hydrodynamic
calculations in the PN limit would still remain important, since
the PN results provide a crucial benchmark
against which future full-GR calculations can be tested.

\begin{figure}
\centering \leavevmode \epsfxsize=7in \epsfbox{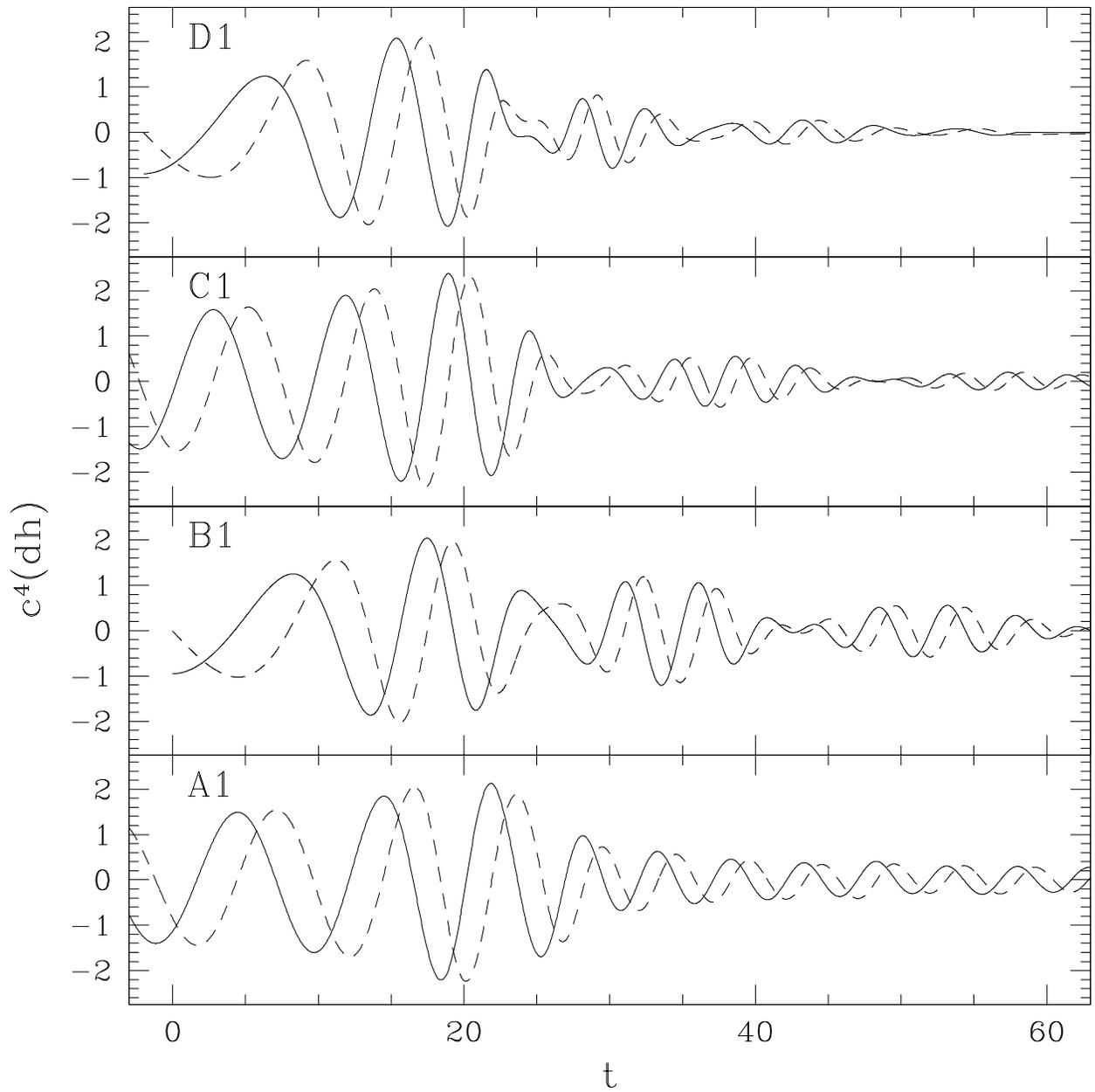}
\caption{Gravity wave signatures for the runs A1 (N, $\Gamma=3$), B1
(PN, $\Gamma=3$), C1 (N, $\Gamma=2$), and D1 (PN, $\Gamma=2$).  All
have $q=1$.  The
solid line shows the $h_+$ polarization, the dashed
line the $h_{\times}$ polarization, both
calculated for an observer at a distance $d$
along the rotation axis.  Note that at $t\protect\gtrsim 60$, 
there is essentially
no gravitational radiation given off by $\Gamma=2$ EOS binaries.}
\label{fig:gwpllong}
\end{figure}

\begin{figure}
\centering \leavevmode \epsfxsize=7in \epsfbox{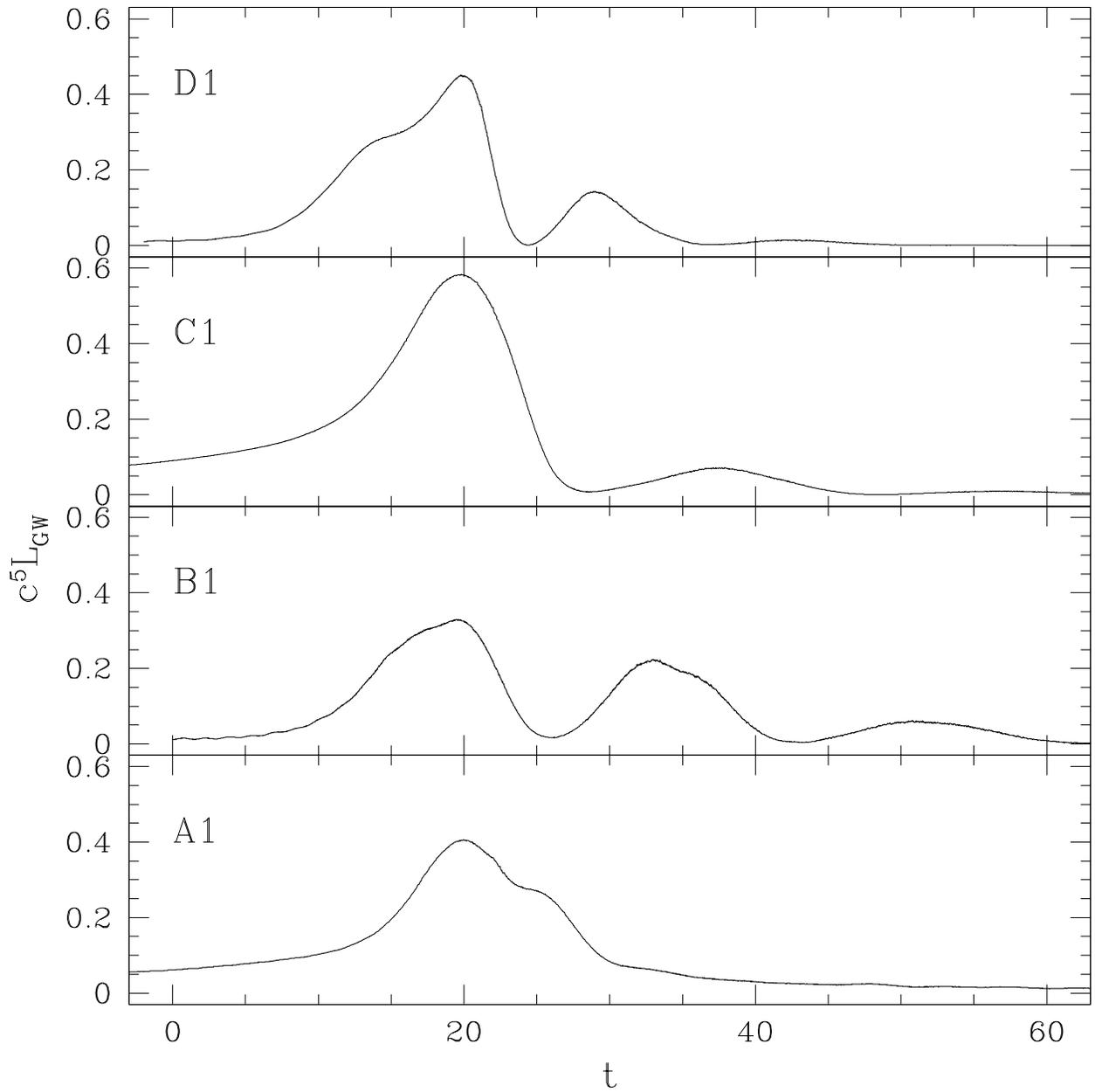}
\caption{Gravity wave luminosity for the same runs as in
Fig.~\protect\ref{fig:gwpllong}.  We see clear evidence for a second gravity
wave luminosity peak in both $\Gamma =2$ runs, but only in 
the PN $\Gamma=3$ run.}
\label{fig:gwlmlong} 
\end{figure}

Gravitational radiation wave forms and luminosities for runs A1, B1, C1, D1 are
shown in Figs.~\ref{fig:gwpllong} and \ref{fig:gwlmlong}.
The two polarizations of gravity waves are
calculated for an observer at a distance $d$ along the rotation axis
of the system, in the quadrupole approximation (see Paper~1, Eqs.~23-25).

As was found for NS with a $\Gamma=3$ EOS (Paper~1), for NS with a 
$\Gamma=2$ EOS PN corrections
serve to lower the maximum gravity wave luminosity.
However, regardless of whether we include 1PN terms, the peak
gravity wave luminosity is larger for the softer EOS.

\begin{figure}
\centering \leavevmode \epsfxsize=7in \epsfbox{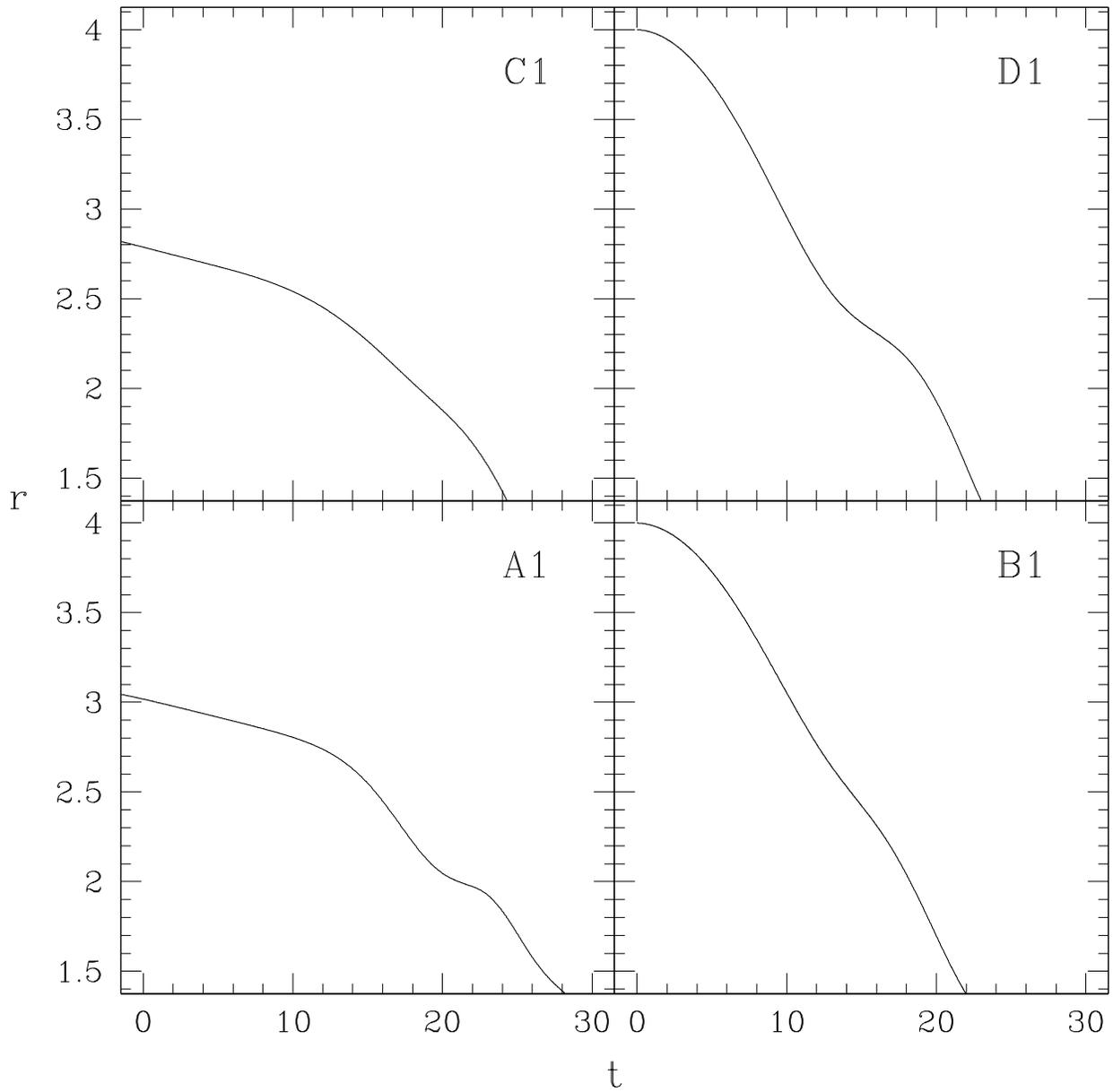}
\caption{Binary separation $r$ for the same runs as in
Fig.~\protect\ref{fig:gwpllong}.}
\label{fig:sep4}.
\end{figure}

Comparing the N and PN runs in Fig.~\ref{fig:gwlmlong}, we see that, besides
differences in the maximum luminosity, there are qualitative
differences in the shape of the first luminosity peak.  Both PN runs
show asymmetry around the peak, with a long rise from $t\simeq10-20$ 
containing a
flatter section at $t\simeq 15-20$, and then a smooth decrease from
$t\simeq 20-25$.  This effect can be attributed to a period of slower binary
infall as the two stellar cores first come into contact (as seen in
Fig.~\ref{fig:sep4}, which shows the binary separation
$r$ as a function of time).  We see that
the effect is more pronounced for the PN run with a $\Gamma=2$ EOS
(D1), which has a
significant drop in the inspiral rate from $t\simeq 14-18$,
corresponding exactly with a similar plateau in the gravity wave luminosity.
The effect is weaker for the PN run with the $\Gamma=3$ EOS (B1),
which displays a more symmetric gravity wave
luminosity peak.

Similar behavior is seen in the N runs.  For the run with a $\Gamma=3$
EOS (A1), we
see a decrease in the inspiral rate from $t=20-25$, which occurs
after the period of maximum gravity wave luminosity, and a
simultaneous plateau in the luminosity immediately after the
peak.  For the $\Gamma=2$ case (C1), we see no apparent decrease in the
inspiral rate, either before or after $t=20$.  Correspondingly, this
was the only run under discussion which showed an almost completely
symmetric luminosity peak with respect to time. 

\begin{figure}
\centering \leavevmode \epsfxsize=7in \epsfbox{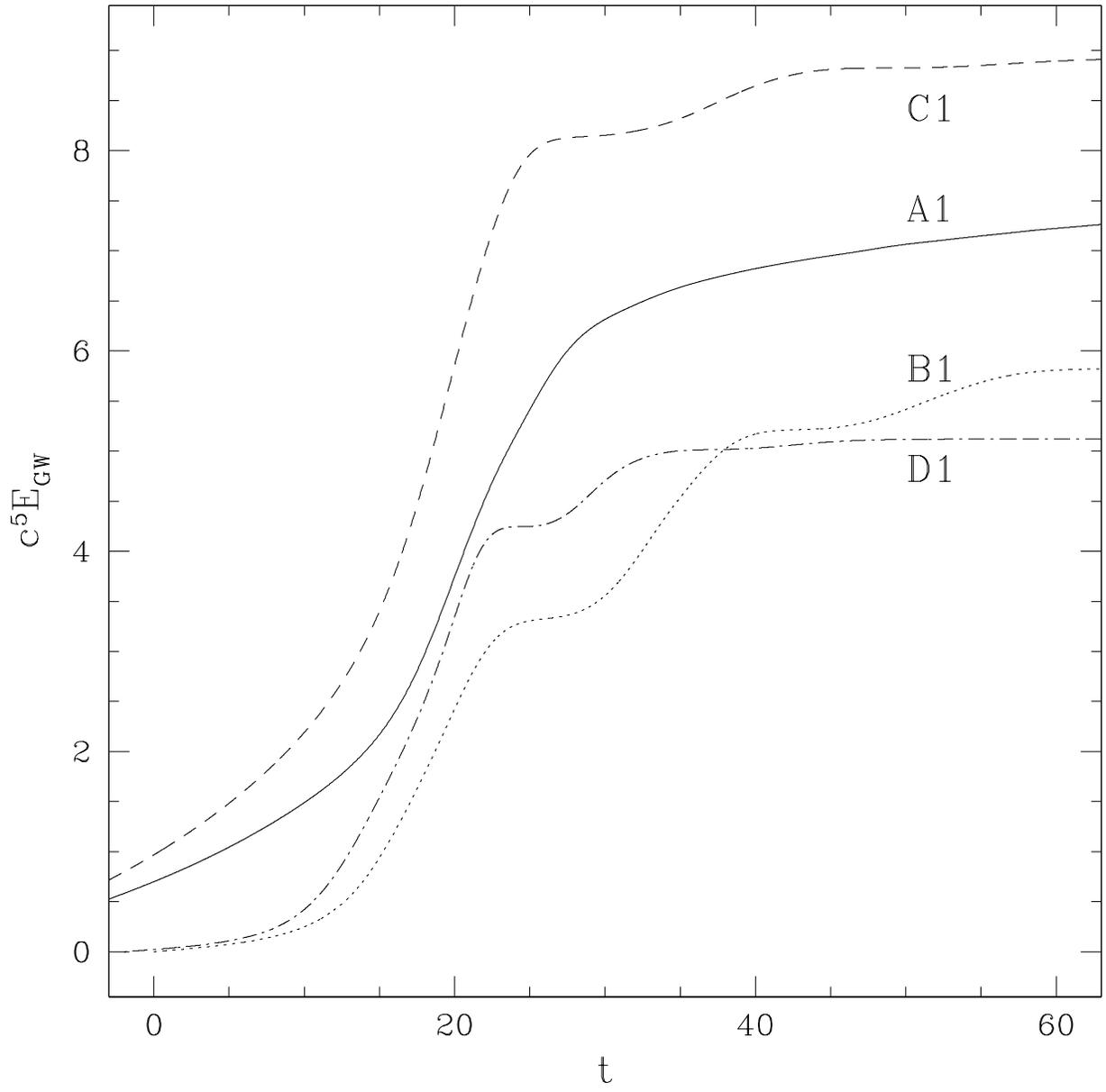}
\caption{Energy lost to gravitational radiation for the same runs shown in
Figs.~\protect\ref{fig:gwpllong}--\protect\ref{fig:sep4}.}
\label{fig:gwen4}
\end{figure}

In Fig.~\ref{fig:gwen4}, we show the energy loss to gravitational
radiation, calculated as the integral of gravity wave luminosity over
time.  We see that the Newtonian runs, A1 and C1, have
higher energy losses, in line with the respective peak gravity wave
luminosities. The secondary peaks in the
$\Gamma=3$ PN run (B1), 
discussed extensively in Paper~1, give a higher net energy
loss than in the $\Gamma=2$ PN run (D1), where more energy is lost
during the first peak (by a significant fraction).
Note that energy losses are measured from the initial point of each
run.  Thus, runs which had negative time offsets have $E_{GW}>0$ at $t=0$.

\begin{figure}
\centering \leavevmode \epsfxsize=7in \epsfbox{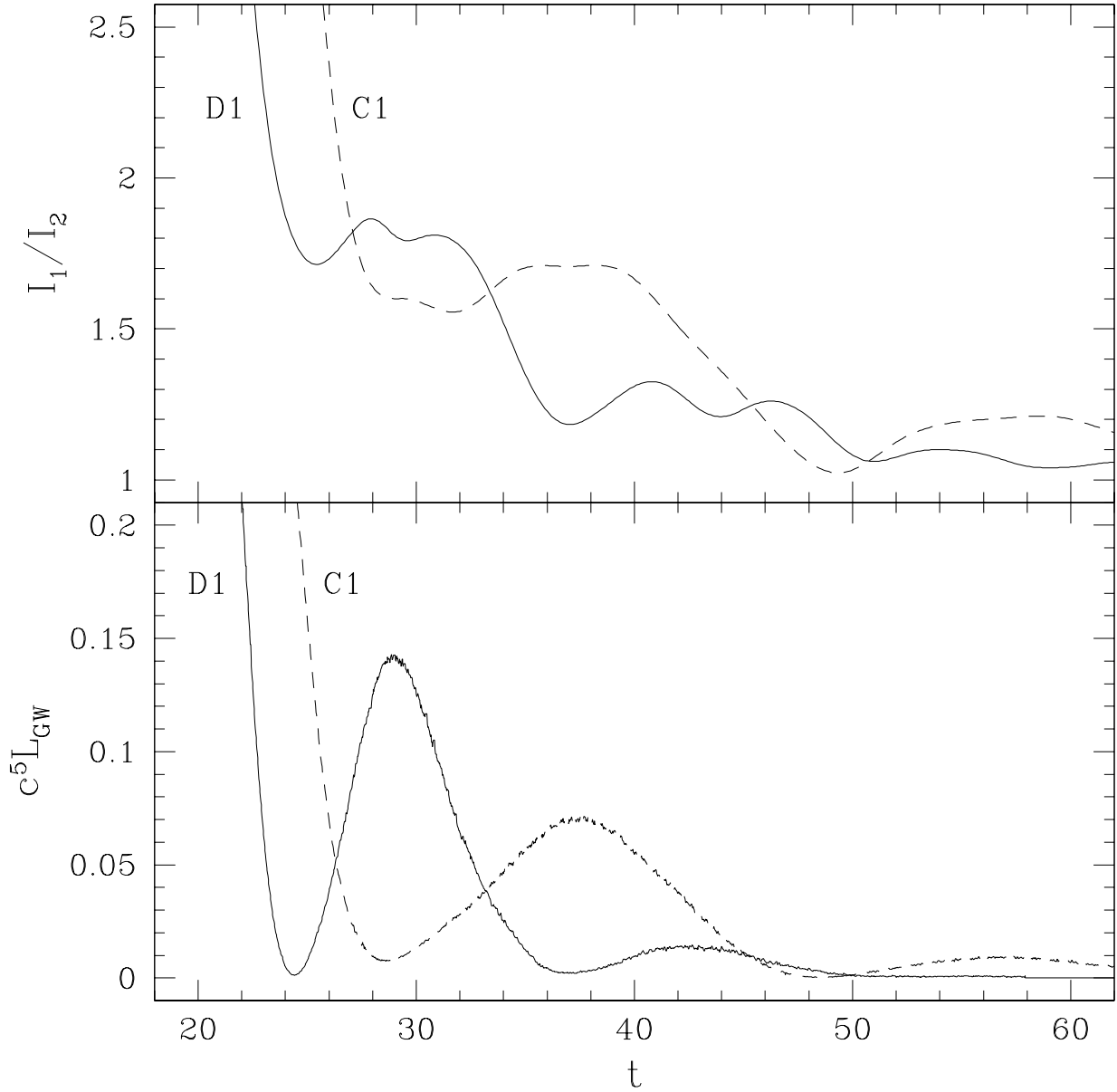}
\caption{Ratio of the principal moments of inertia in the
equatorial plane for the remnants of runs C1 (N, $\Gamma=2$; dashed
lines) and D1 (PN, $\Gamma=2$; solid lines), 
compared to the corresponding gravity wave luminosities at late times.
We see a clear correlation between the two quantities, 
as the remnants relax towards an axisymmetric, oblate
configuration.  
Here the remnants are defined by the density cut $r_*>0.04$,
which includes the entire binary initially, but only the inner part
of the merger at later times.}
\label{fig:mom2} 
\end{figure}

More surprisingly, perhaps, we note that a second luminosity peak is visible
not only in the $\Gamma=2$ PN 
run (D1), but also in run C1, for a $\Gamma=2$ Newtonian coalescence.  In
Paper~1, it was shown that the gravity wave luminosity is extremely well
correlated with the ratio of the first and second principal moments of
inertia, $I_1$ and $I_2$, 
in effect giving a measure of the ellipticity of the remnant
in the orbital plane.  A similar analysis is shown in
Fig.~\ref{fig:mom2}, which compares runs C1 and D1.
Note that we have used a higher density cut for this plot 
than was used for the $\Gamma=3$
remnants in Paper~1.  Here only, we define the inner remnant to consist of all
SPH particles with local densities $r_*>0.04$.
We see again a strong correlation between
ellipticity and gravity wave luminosity. In run D1, the
small-amplitude oscillations with period $T\simeq 5$ are caused by an
interaction with the outer material of the remnant.  If we lower our
density cut to include all SPH particles with local densities
$r_*>0.005$ (as in Paper~1),
thereby including more of the tenuous material outside of the core of
the remnant, the moment of inertia ratio shows only this oscillation,
which damps out over time.  Since this material contributes only
weakly to the quadrupole moment, we conclude that 
it is the dynamics inside the core that controls the gravity wave signal. 

Comparing the results of our run C1, which includes the effects
of radiation reaction, with the completely Newtonian $\Gamma=2$
run shown in RS2, we conclude that the asymmetry induced by the larger
tidal lag angle is responsible for
the existence of a second gravity wave luminosity peak, even for systems that
eventually reach oblate, non-radiating configurations.  To confirm
this, we checked our results against a  completely Newtonian
$\Gamma=2$ test run (without radiation reaction terms, and starting from
$r_0=2.7$), and found that a second peak is indeed absent in this case
(in agreement with RS2).

Careful inspection of Fig.~7 of Paper~1 (where run A1
is referred to as the N run), reveals very small irregularities
in the otherwise smooth
gravity wave luminosity at $t\simeq33$ and~48, which correspond to
the secondary gravity wave luminosity peaks
found in run B1. Clearly, the existence of secondary luminosity peaks
seems to be reasonably universal in these simulations.  For run A1, however,
modulation of the moment of inertia ratio is virtually 
absent, so that the effect is extremely small in this case.

\subsection{Dependence on the binary mass ratio}

The dependence of the peak amplitude 
$h_{\rm max}\equiv (h_+^2+h_{\times}^2)_{\rm max}^{1/2}$ 
of gravitational 
waves on the mass ratio $q$ appears to be very strong.
In RS2, an approximate power law $h_{\rm max}\propto q^2$ was derived
for nearly equal-mass systems, 
on the basis of two purely Newtonian calculations for $q=1$ and $q=0.85$.
This is considerably steeper than the na\"{\i}ve scaling obtained for 
two point masses in a Keplerian orbit, which gives $h_{\rm max}\propto q$. 
Such a linear scaling is obeyed (only approximately, because of
finite-size effects) by the wave amplitudes of the various systems
{\it prior\/} to final coalescence.
For determining the maximum amplitude during the merger, however, hydrodynamics
must be taken into account. In a system with $q\ne 1$, the more massive
star tends to play a far less active role in the hydrodynamics
and, as a result, there is a rapid suppression of the
radiation efficiency as $q$ departs even slightly from unity.
For the peak luminosity of gravitational radiation RS found
approximately $L_{\rm max}\propto q^6$. Again, this is a much steeper 
dependence than
one would expect based on a simple point-mass estimate, which gives
$L\propto q^2(1+q)$. The results of RS were all for
initially synchronized binaries, but very similar results have been
obtained by Zhuge et al.\ \cite{Zhu2} for binaries containing initially 
nonspinning stars with unequal masses. 

\begin{figure}
\centering \leavevmode \epsfxsize=6in \epsfbox{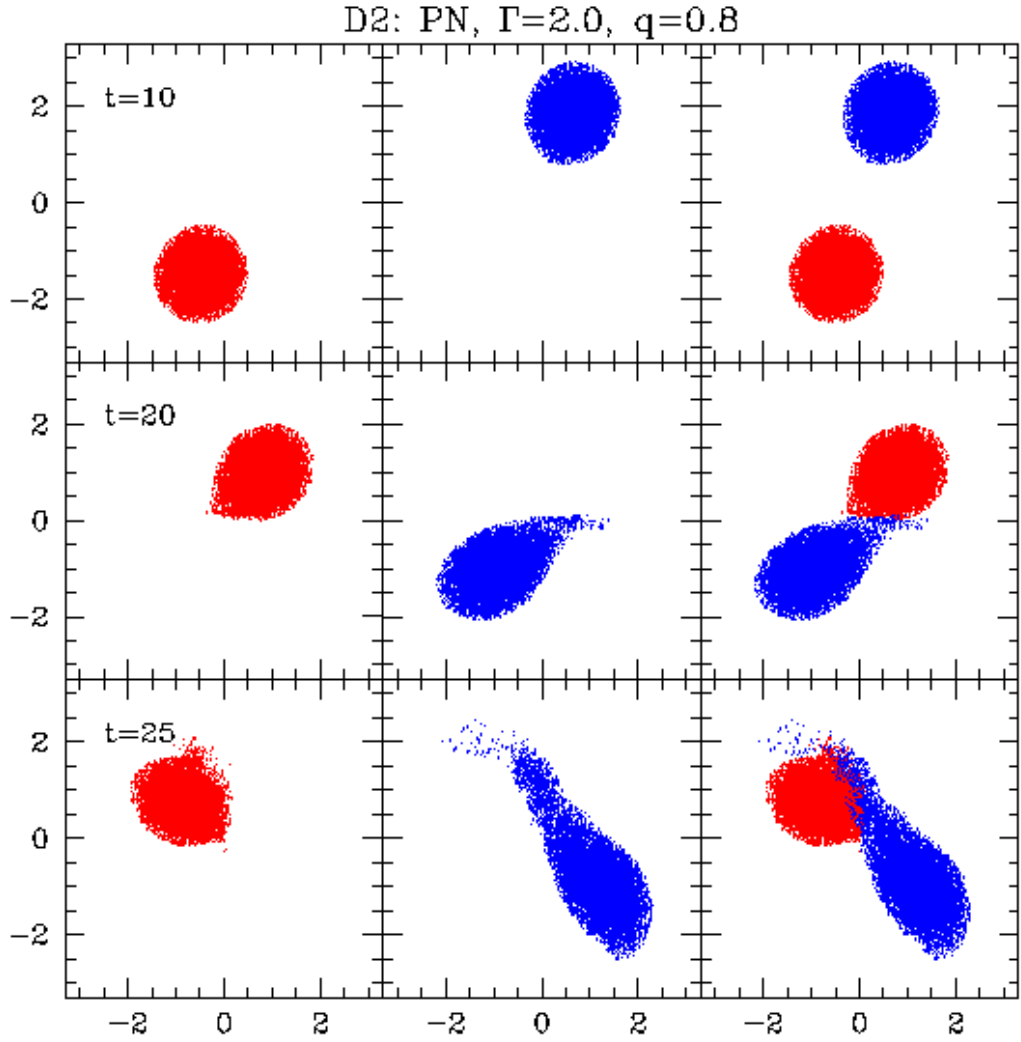}
\caption{Evolution of the system for run D2 with $\Gamma=2$,
$q=0.8$, and 1PN corrections included.  Projections of a
random subset of $20\%$ of all SPH particles onto the orbital (x-y)
plane are shown at various times.  Panels on the left show the
primary, center panels show the secondary, and panels on the right the
entire system.}
\label{fig:xy2p08a}
\end{figure}

\begin{figure}
\centering \leavevmode \epsfxsize=6in \epsfbox{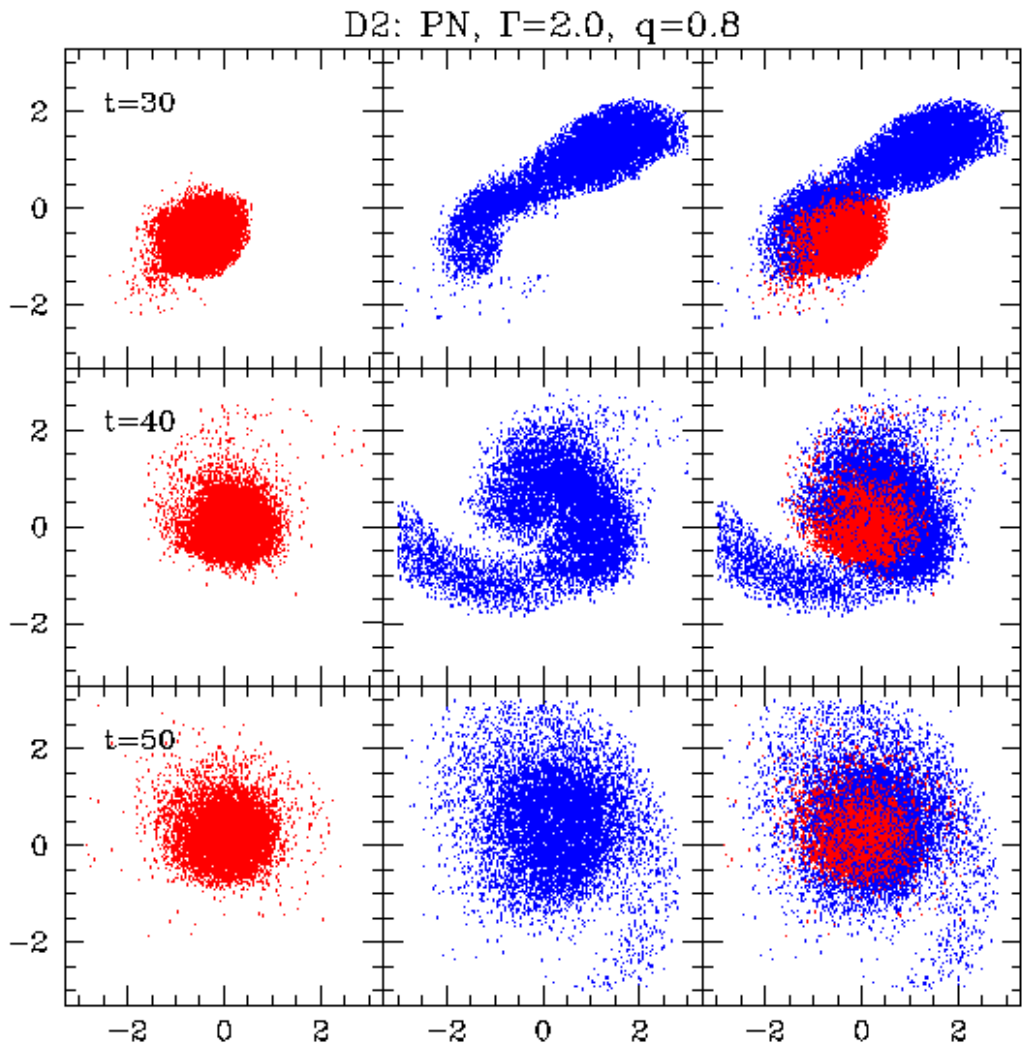}
\caption{Evolution of run D2, continued.}
\label{fig:xy2p08b}
\end{figure}

The role of the primary and secondary 
is shown qualitatively in Figs.~\ref{fig:xy2p08a} and
\ref{fig:xy2p08b} for run D2,
which uses a $\Gamma=2$ EOS with $q=0.8$, and included 1PN
terms.  The panels on the left
show the primary, center panels the secondary, and those on the right
the combined system.
We see that the evolution is markedly different from that
of the $q=1$ binary shown in Fig.~\ref{fig:xy2p}.  Here, there is a
single spiral arm, which is formed from the secondary as it gets
tidally disrupted.  Shortly
after first contact is made, a stream of matter flows from the
secondary toward the primary, landing on the trailing side (orbital rotation
is counterclockwise), as a result of the orbital motion and the significant
tidal lags present in the system upon contact.  As the coalescence proceeds,
the secondary is tidally stretched, with the outer portion spun
out of the system while the inner part is accreted by the primary.
Throughout this evolution, the primary remains relatively undisturbed,
except for a small layer near its surface.  By $t\simeq 40$, mass shedding
is triggered and the system develops a single spiral
arm, composed entirely of matter from the secondary as it is completely
disrupted.  Note that this single spiral arm is even wider than those seen 
in the $q=1$ coalescence (with a width comparable to the
initial NS radius).  There is also an extremely small amount of mass shedding from
the outer edge of the primary, where it joins with a high-velocity stream of
matter from the secondary, on the side opposite the single spiral arm.
Finally, by $t\simeq 50$, what was once the core of the secondary has
fallen onto the primary, and the spiral arm has begun to dissipate,
forming a low-mass halo around the system.

\begin{figure}
\centering \leavevmode \epsfxsize=7in \epsfbox{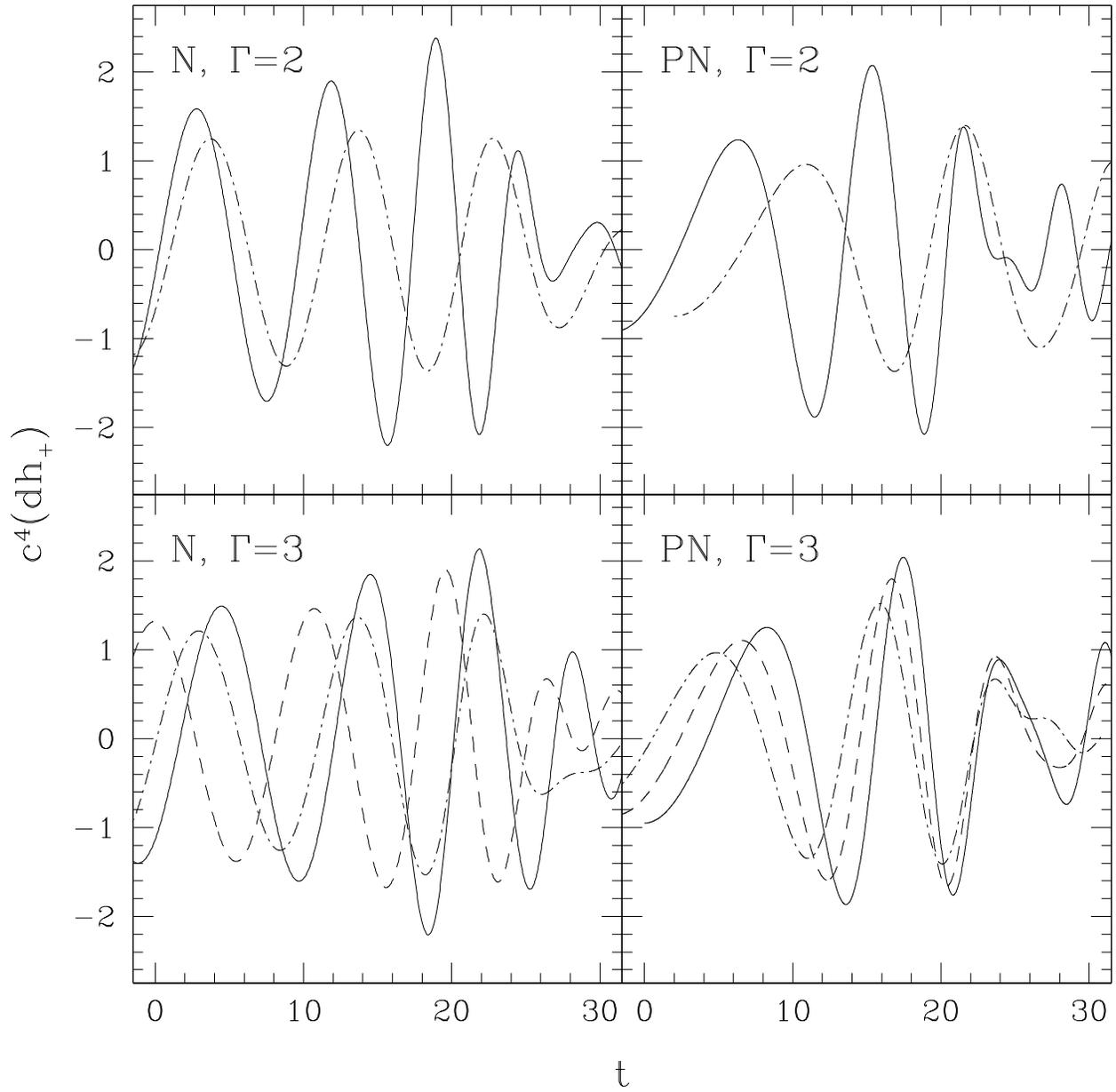}
\caption{Gravity wave amplitude $h_+$ for an observer
located at a distance $d$ along the rotation axis, comparing systems
with different mass ratios.  The solid lines correspond to
$q=1$, the dashed lines to $q=0.9$, and the dot-dashed lines to $q=0.8$.}
\label{fig:gwpl4}
\end{figure}

\begin{figure}
\centering \leavevmode \epsfxsize=7in \epsfbox{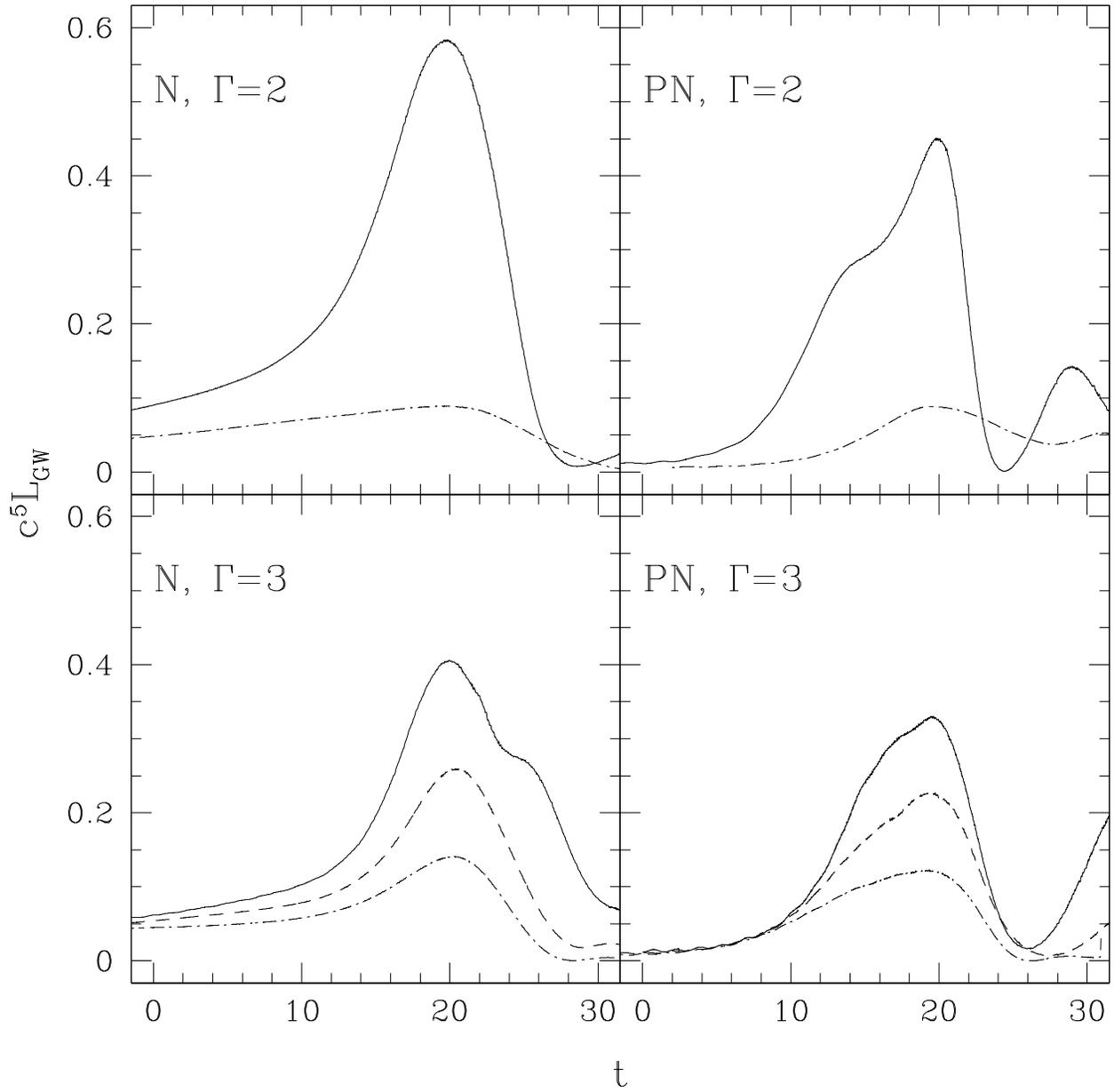}
\caption{Gravity wave luminosity for the same runs shown in
Fig.~\protect\ref{fig:gwpl4}.  Conventions are in
Fig.~\protect\ref{fig:gwpl4}.}
\label{fig:gwlm4}
\end{figure}

In Figs.~\ref{fig:gwpl4} and \ref{fig:gwlm4}, we show the gravitational radiation
wave forms and luminosity for all but two of the synchronized
binary simulations.  For clarity, we only show results for runs
A1, A3, and A5 in the $\Gamma=3$ N plot, since the other runs can be
safely interpolated from those present.  We note that for the
$\Gamma=3$ EOS, the morphology of the peaks seen in
Fig.~\ref{fig:gwlmlong} seems to be present for all mass ratios.
We find plateaus in the gravity wave luminosity before the peak
for PN runs, and after the peak for N runs.  This behavior is much
harder to see in the $\Gamma=2$ runs with q=0.8, which have lower peak
luminosities in general than their $\Gamma=3$ counterparts.

\begin{figure}
\centering \leavevmode \epsfxsize=7in \epsfbox{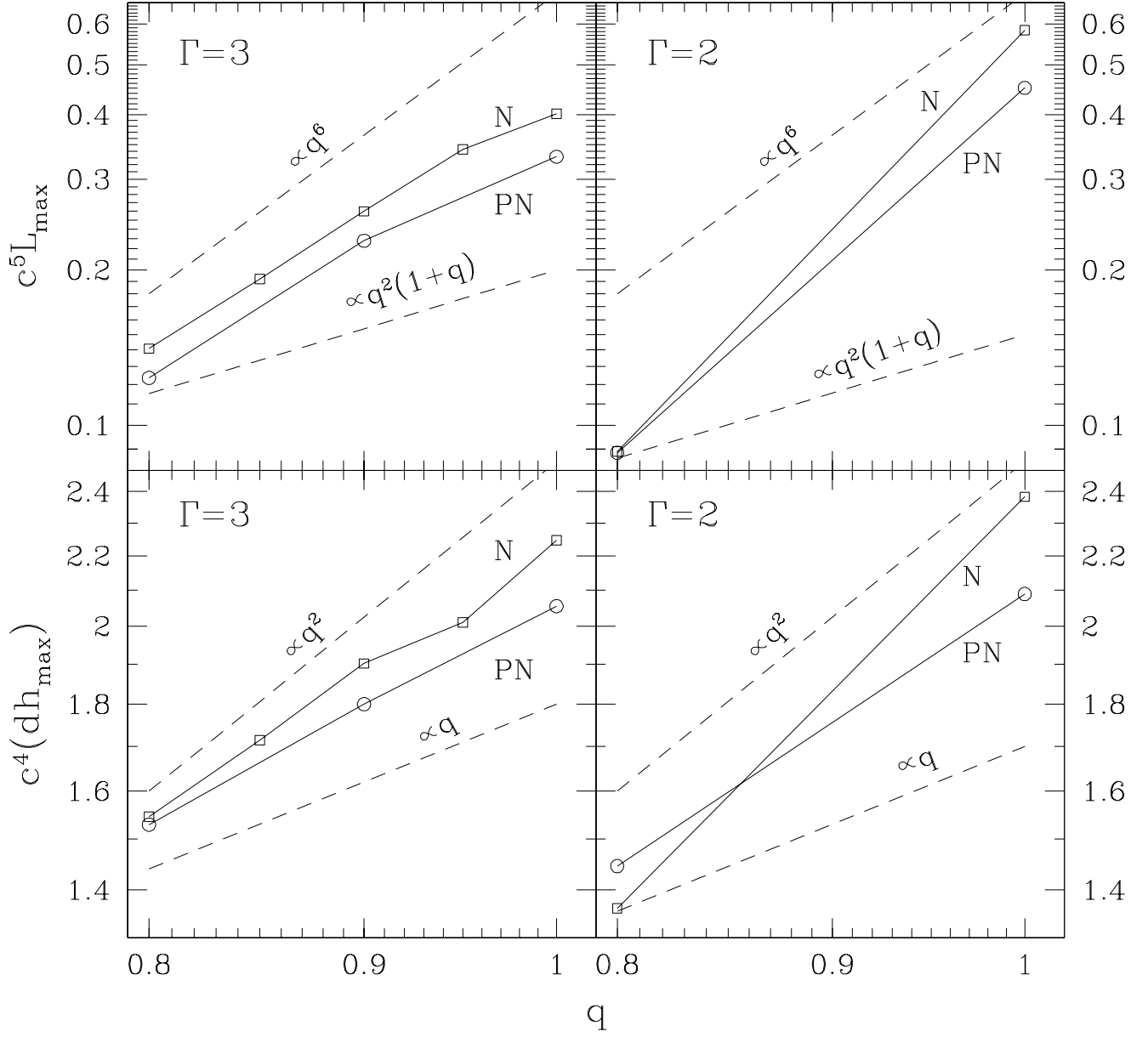}
\caption{Dependence of the maximum gravity wave amplitude and luminosity
on the mass ratio (for all the synchronized binaries).
The Keplerian point mass approximation
gives $h_{\rm max}\propto q$ and $L_{\rm max}\propto q^2(1+q)$.
Instead, the strictly Newtonian hydrodynamic calculations of RS2 give
approximate power laws 
$h_{\rm max}\propto q^2$ and $L_{\rm max}\propto q^6$ 
for nearly equal-mass binaries containing $\Gamma=3$ polytropes.
Note that both axes are plotted logarithmically.} 
\label{fig:4vsq}
\end{figure}

In Fig.~\ref{fig:4vsq}
we show the maximum gravity wave strain and luminosity
for $\Gamma=3$ and $\Gamma=2$
EOS binaries, plotted as a function of the mass ratio.  
We find in all cases that
the power-law dependence is steeper than would be predicted by the
point-mass approximation.  For the gravity wave strain, we see 
a slightly steeper power law for N runs than for PN
runs.  For $q=1$, N runs have a higher peak strain, but, for both
EOS, the $q=0.8$ binaries show a higher peak strain in the PN case
than in the N case. We conclude that the strictly Newtonian scaling
obtained by RS2, $h_{\rm max}\propto q^2$, 
remains approximately valid
for Newtonian binaries with radiation reaction effects, but is almost
certainly steeper than the dependence that will be present in realistic
NS binaries (based on our results including 1PN effects).

For the peak gravity wave luminosity, we see a much stronger
dependence on the EOS.  For the softer, $\Gamma=2$ EOS, we see that the
strictly Newtonian RS2 fit of 
$L_{\rm max}\propto q^6$ (obtained for $\Gamma=3$ polytropes!) 
is not nearly as  steep as
the correct relation.  Additionally, the difference
between N and PN runs in magnitude is less pronounced than for
the $\Gamma=3$ EOS, where we have a $\sim 20\%$ difference
rather independent of the mass ratio.  For $\Gamma=3$, we see that
the RS2 fit is slightly too steep, although the correct power law 
remains steeper than the point-mass approximation.

\subsection{Irrotational initial condition}\label{sec:irrotres}

\begin{figure}
\centering \leavevmode \epsfxsize=6in \epsfbox{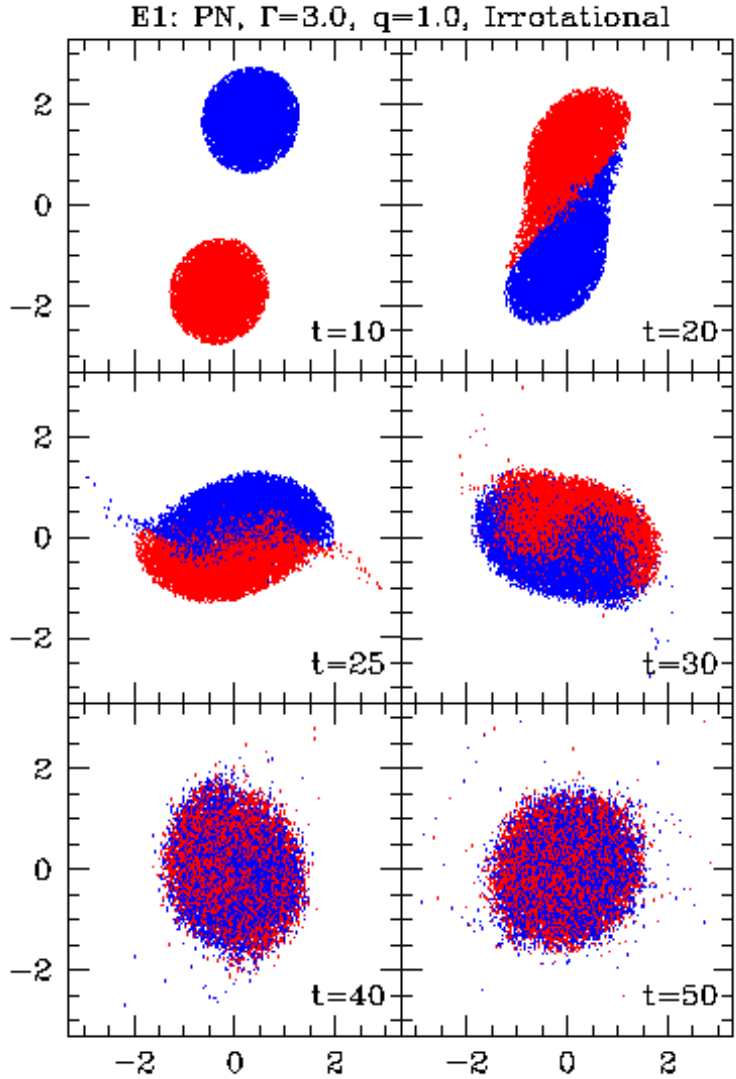}
\caption{Evolution of the $\Gamma=3$, $q=1$ irrotational binary system
(E1).
Conventions are as in Fig.~\protect\ref{fig:xy2n}.  
We see evidence for spiral arms
forming after the initial merger, but containing much less mass than
in the synchronized case.  Note also that a
longer time is required after merger to relax to a triaxial configuration.}
\label{fig:xyirr}
\end{figure}

To study the effects of the initial spins on the evolution of the
merger we
compare our run E1, with the irrotational initial condition described in
Section~\ref{sec:irrot}, with run B1, the PN run described extensively
in Paper~1, which started from a synchronized initial condition.  Particle
plots for run E1 are shown in Fig.~\ref{fig:xyirr}.  Comparing to
Fig.~2 of Paper~1, we see
that in both cases, the stars develop a large tidal lag immediately before
merger, leading to an ``off-center'' collision and a highly
asymmetrical merger until $t\simeq 50$.  A major
difference, however, is the absence of mass shedding as
the irrotational NS first make contact before $t=20$.  We see a very small
amount of mass shedding and thin spiral
arm formation from $t\simeq 25-35$, involving a very small fraction of the
mass that was deposited into the outer halo 
for the synchronized case. 
Note that mixing of the fluids originating from the two different stars 
does not occur immediately after contact. Instead, as the
two NS surfaces come into contact with opposite tangential velocities (in
the corotating frame), a long vortex sheet forms at the interface between
the two fluids (around $t\simeq 20$).
This vortex sheet is Kelvin-Helmholtz unstable 
and breaks into a turbulent layer that can then gradually mix
the fluids \cite{RS94}. 
Unfortunately, the spurious shear viscosity inherent in
any discretized simulation is much larger than the true viscosity of
the NS fluids, and therefore the evolution in the turbulent region is
very likely to be dominated by resolution-dependent numerical effects.

Noting that the orbital rotation for the binary is
counter-clockwise, we see that all mass shedding into the spiral arms actually
occurs from the leading edge of each star.  Particles that were
initially located at small radius relative to the center of mass of
the system travel along the vortex sheet between the two stars before
being spun off.  This is in direct contrast to the synchronized case,
in which mass shedding occurs almost entirely from regions that are
initially at large radius, on the outside of each star. Material there is
ejected from the merging binary system as it spins up and
overflows its effective potential well through the outer Lagrange points.  
Here we find no mass shedding via this mechanism, in agreement
with previous work on irrotational systems \cite{RJS}.

\begin{figure}
\centering \leavevmode \epsfxsize=6in \epsfbox{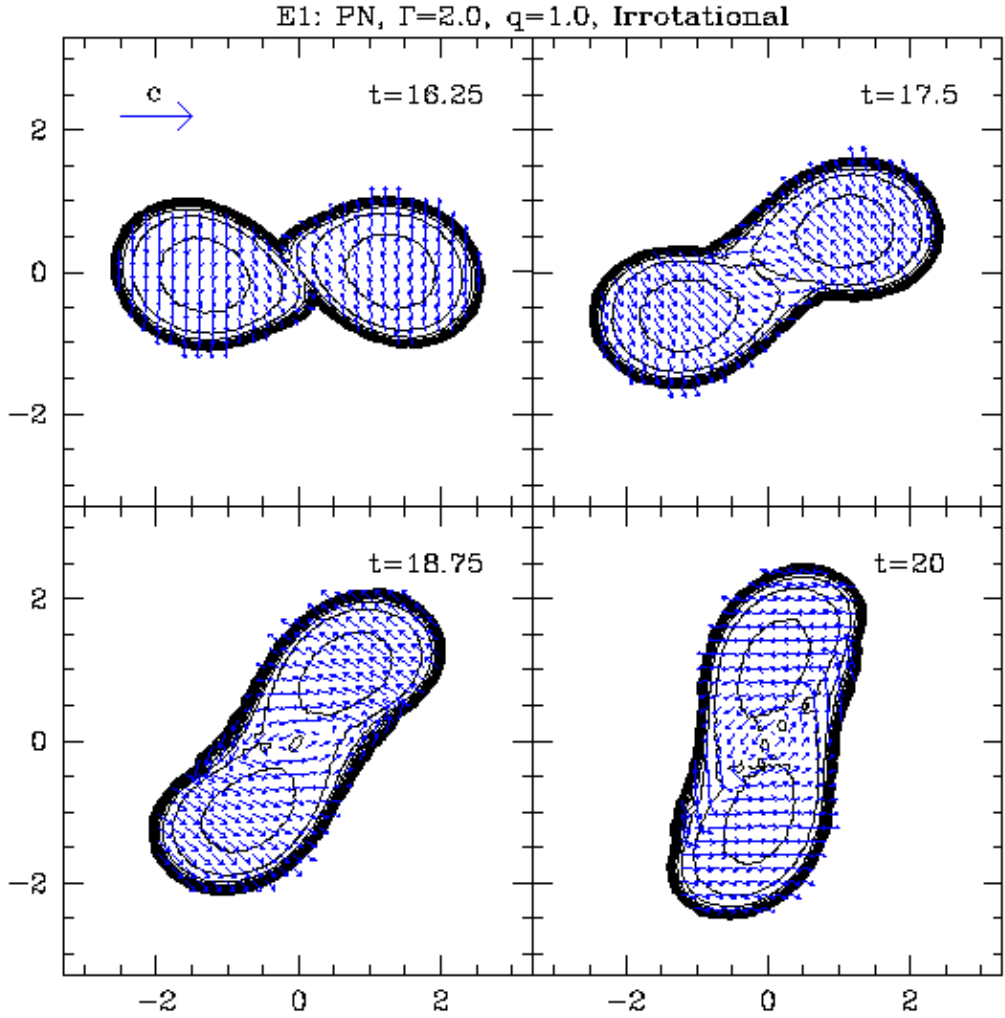}
\caption{Density contours and velocity field for the irrotational system
in run E1.
Density contours are spaced logarithmically, 4 per decade. 
The velocity field is shown in the {\it inertial frame.\/}  A vector
representing the speed of light ($c=c_{2.5PN}$, the physical value) is
shown for comparison.}
\label{fig:dvirr}
\end{figure}

\begin{figure}
\centering \leavevmode \epsfxsize=6in \epsfbox{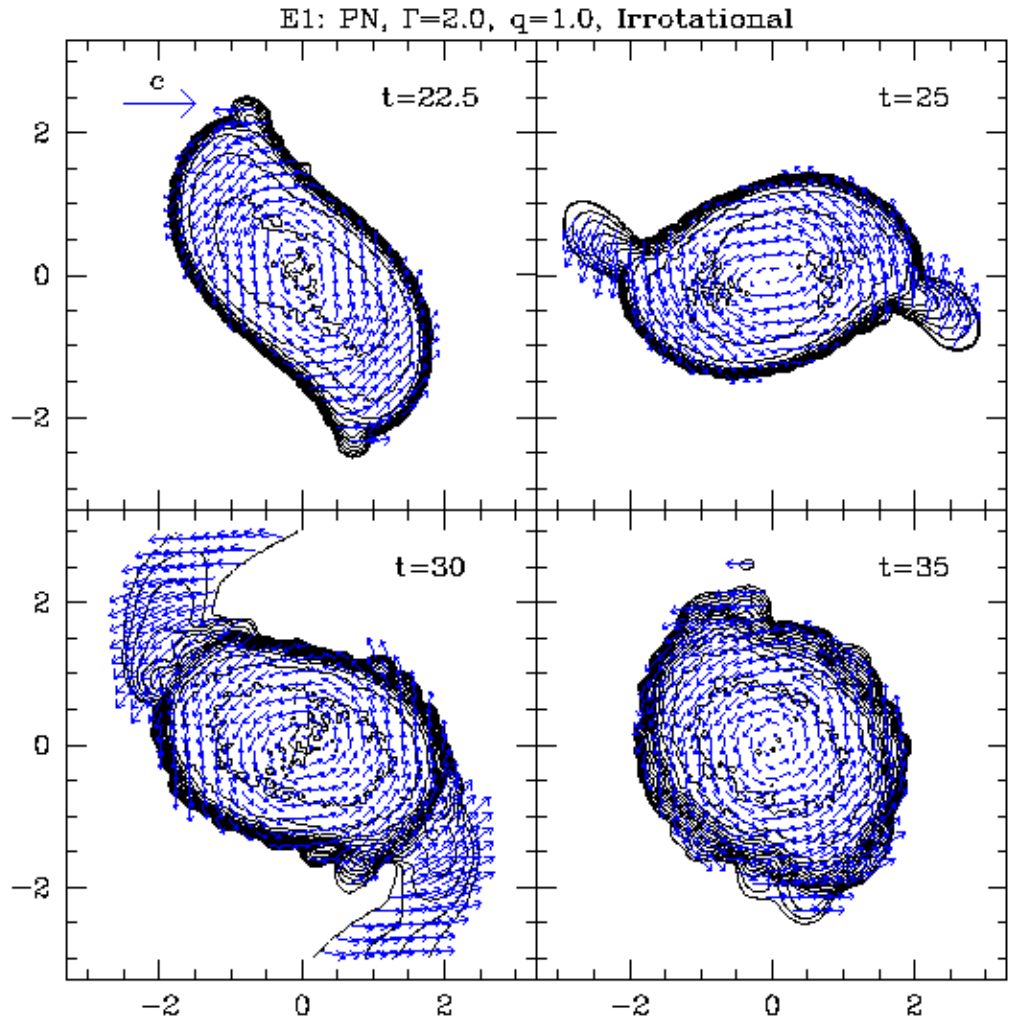}
\caption{Density contours and velocity field for run E1, continued.}
\label{fig:dvirr2}
\end{figure}

\begin{figure}
\centering \leavevmode \epsfxsize=6in \epsfbox{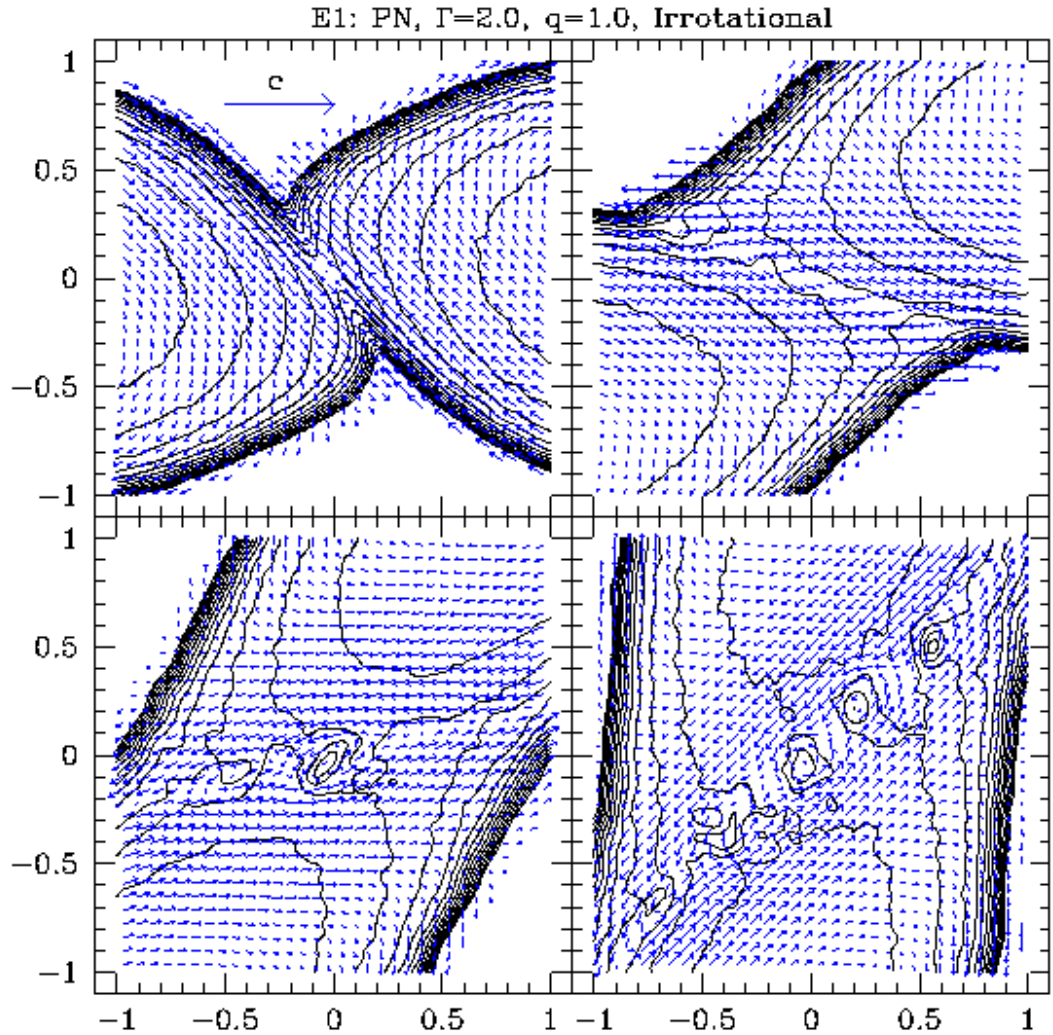}
\caption{Density contours and velocity field for the irrotational 
system in run E1, here
focusing on the inner region of the merger.  Panels correspond to the
same times as shown in Fig.~\protect\ref{fig:dvirr}.
Density contours are spaced logarithmically, 8 per decade. 
The velocity field is shown in the {\it corotating frame \/}, as defined by
Eq.~\protect\ref{eq:omcorot}.  As the
surfaces of the two stars come into contact, 
they form an vortex sheet which extends over the full
length of the interface.}
\label{fig:dvirrz}
\end{figure}

\begin{figure}
\centering \leavevmode \epsfxsize=6in \epsfbox{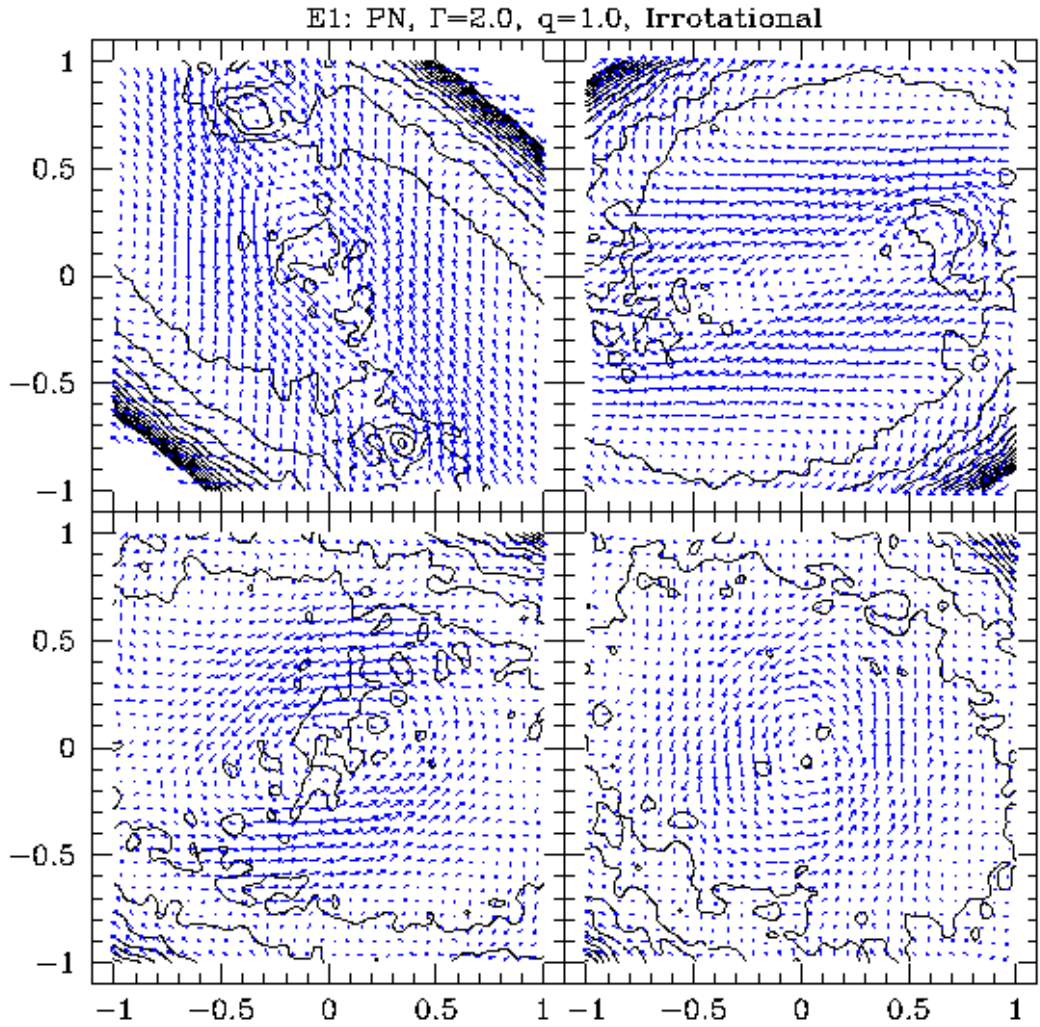}
\caption{Density contours and velocity field in the inner region for
run E1, continued. Panels correspond to the same times as in Fig.~15.}
\label{fig:dvirrz2}
\end{figure}

In Figs.~\ref{fig:dvirr} and \ref{fig:dvirr2} 
we show density contours during the merger of
run E1, with the velocity field of the fluid in the inertial
frame overlaid. Mass shedding and spiral arm formation starts at $t\simeq22$.  
In Figs.~\ref{fig:dvirrz} and~\ref{fig:dvirrz2} 
we show the
density and velocity field of the inner regions of the merger, with
the velocities shown with respect to the mean corotating
 frame of the entire system. Specifically, we define the mean
corotating angular velocity $\Omega_c$ by taking a weighted average over
the tangential velocities of all SPH particles, 
\begin{equation}
\Omega_c\equiv \frac{\sum_i m_i[(xv_y-yv_x)/r_{\rm cyl}]_i}
{\sum_i m_i (r_{\rm cyl})_i},
\label{eq:omcorot}
\end{equation}
where the cylindrical radius is defined as $r_{\rm cyl}\equiv
(x^2+y^2)^{1/2}$.
It is easy to see that this definition reduces to the correct orbital
angular velocity of the initial binary system, and to the correct
angular velocity of any single, rigidly rotating object. We see
that the vortex sheet formed during the merger lasts from the point of
first contact at $t\simeq 15$ until $t\simeq 22$.  As the
coalescence proceeds, turbulence develops along the sheet, with
vortices appearing immediately at the very center of the remnant, then
at several points further out.  We note that the central vortex
is initially spinning in the direction of the orbital motion, as the
initial misalignment causes matter to fall in around the center.  The
outer vortices are concentrated around regions of higher density, when
matter from deeper inside each NS continues to fall onto the vortex
sheet.  From $t\simeq 22$ to $t\simeq 25$, a slightly different pattern
emerges.  As the core regions of the original NS continue to approach
each other, the velocity field between the cores remains turbulent, but a
coherent velocity field develops around the cores.  Counterspinning
material located toward the outer parts of the remnant connects and forms
a simple pattern of circulation. Here
the angular velocity appears to become negative as we move
away from the center (in the inertial frame, this represents the
angular velocity decreasing with increasing distance from the rotation axis).  
Material in the
center, still somewhat turbulent but retaining the original corotating
pattern, has much less angular momentum, leading to a pattern of
strongly differential rotation throughout the merger remnant (See
Sec.~\ref{sec:remnant}).

\begin{figure}
\centering \leavevmode \epsfxsize=7in \epsfbox{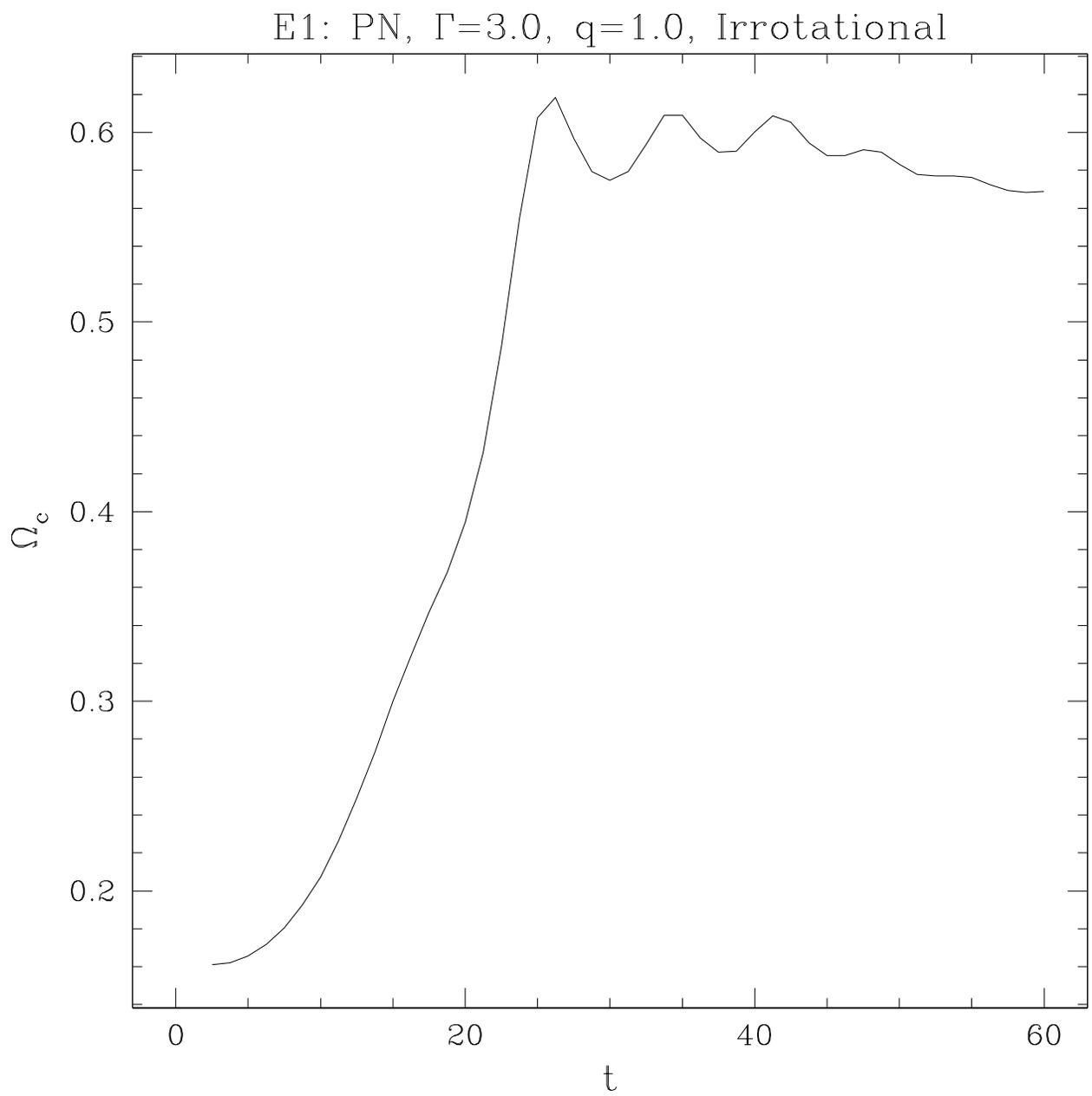}
\caption{Evolution of the mean angular velocity of the corotating frame $\Omega_c$
for run E1 (as
defined by Eq.~\protect\ref{eq:omcorot}).}
\label{fig:omega}
\end{figure}

The evolution of the mean corotating angular velocity is shown in
Fig.~\ref{fig:omega}.  It rises monotonically during the initial coalescence
until $t\simeq 25$, when the inner NS cores merge.
Later it fluctuates with a small amplitude around a nearly constant 
maximum value
$\Omega_c\simeq 0.6 \simeq 1.4\times10^4\,{\rm rad}\,{\rm s}^{-1} 
(M/1.5\,M_\odot)^{1/2} (R/10\,{\rm km})^{-3/2}$
as the remnant contracts and relaxes. By $t\simeq50$ it 
settles down with a slowly and steadily 
decreasing trend as the final triaxial configuration continues to lose
angular momentum to gravitational radiation.

\begin{figure}
\centering \leavevmode \epsfxsize=7in \epsfbox{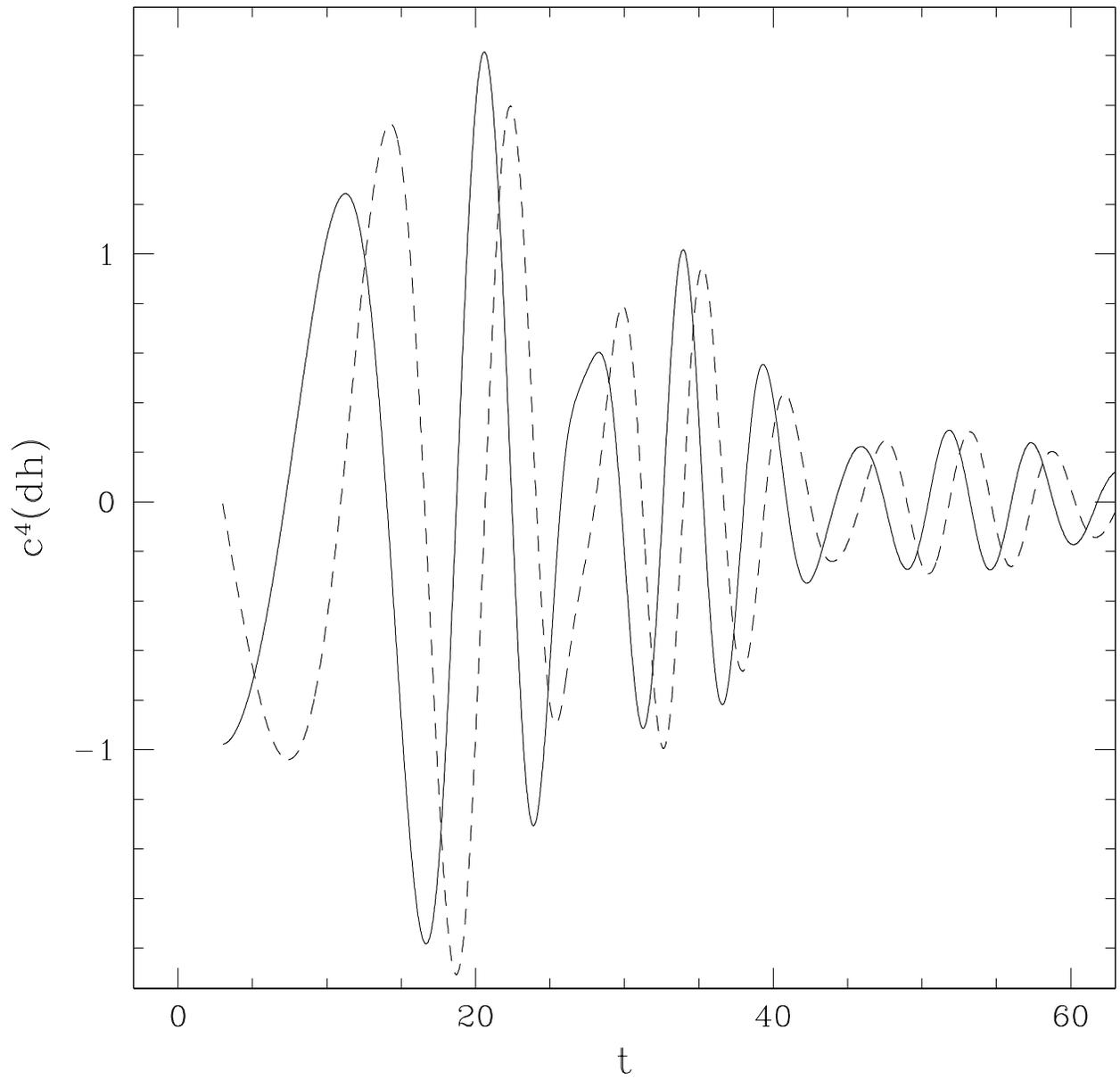}
\caption{Gravity wave signal for the run E1 (irrotational).  
Conventions are as in Fig.~\protect\ref{fig:gwpllong}.}
\label{fig:gwplirrot}
\end{figure}

\begin{figure}
\centering \leavevmode \epsfxsize=7in \epsfbox{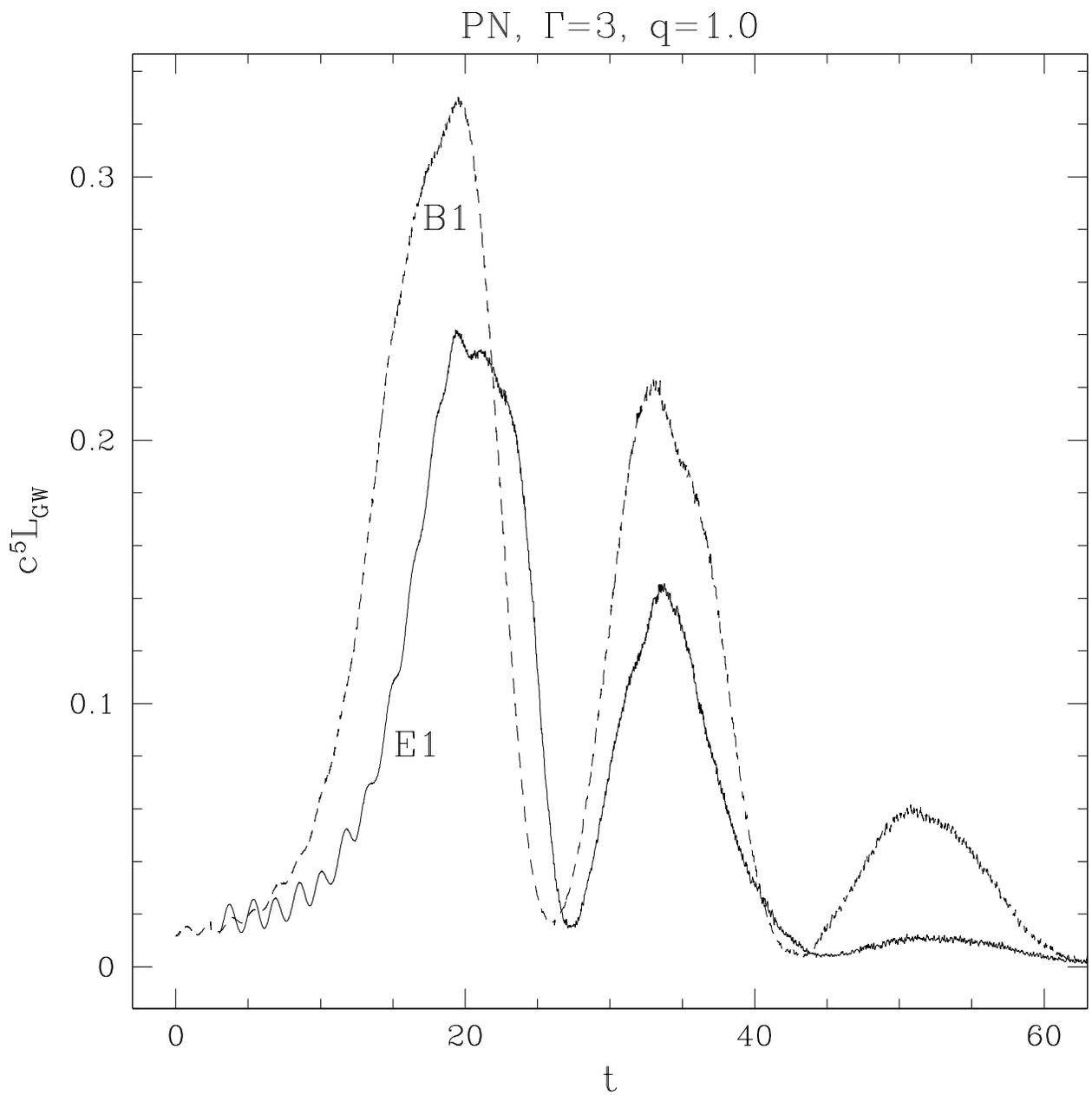}
\caption{Gravity wave luminosity for run E1 (irrotational)
compared with that of run B1 (with a synchronized initial condition).
Both runs include 1PN corrections, and have $\Gamma=3$ and $q=1$.}
\label{fig:gwlmirrot}
\end{figure}

The gravitational radiation wave form for run E1 is shown in Fig.~\ref{fig:gwplirrot}.
There are no major qualitative differences with the wave form obtained for
run B1, which has the same EOS and mass ratio, but
a synchronized initial condition (see Fig.~3).
The gravity wave luminosity is shown in Fig.~\ref{fig:gwlmirrot},
along with that of run B1.
We see that irrotational initial conditions lead to a slight reduction in
the maximum gravity wave luminosity, simply because the system has a smaller total 
angular momentum. The small-amplitude oscillations in the luminosity of run E1
for $t\simeq0-15$
result from small deviations from quasi-equilibrium in the initial binary
of magnitude $\delta\rho_c/\rho_c\simeq 0.05$.
This small amplitude gives us confidence that the use of self-similar PN
ellipsoids in the construction of our irrotational initial condition
represents a very small source of error for these
calculations. Our irrotational ellipsoids, although still approximate,
are considerably better than the spherical models used in all previous
studies (see, e.g., \cite{Zhu2,RJS}).  

Both the E1 and B1 runs show a
significant second luminosity peak in the gravity wave emission, although the
magnitude of the second peak is again smaller for the irrotational case.  
This second peak
occurs at $t\simeq33$ for both runs, indicating that the period between
peaks in the gravity wave signal seems to
depend primarily on the NS EOS and not on the initial spins.  
Note also in Table~\ref{table:runs} that the
second peaks for runs with a $\Gamma=2$ EOS are at different times,
whereas for $\Gamma=3$, the peaks are always coincident.  A third peak
is also visible for the irrotational run, but it is of much lower amplitude,
indicating that the final remnant is more nearly axisymmetric.

\subsection{Structure of the final merger remnants}\label{sec:remnant}

\begin{figure}
\centering \leavevmode \epsfxsize=7in \epsfbox{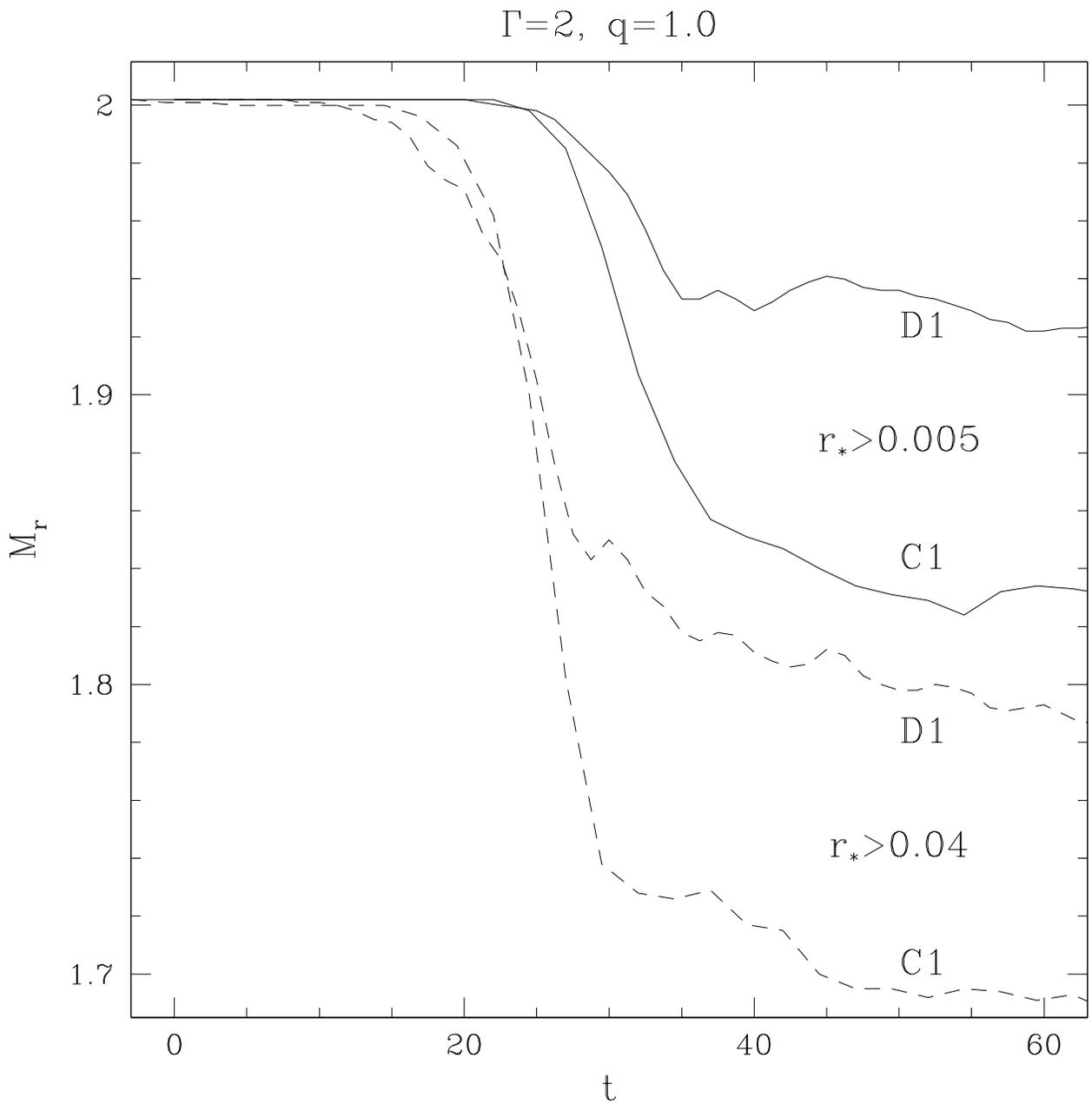}
\caption{Evolution of the remnant mass for the $\Gamma=2$, $q=1$
runs (C1 and D1).  Suppression of mass shedding in the PN case
leads to a higher final remnant mass.  The density cut $r_*>0.04$, shown as
dashed lines, includes the inner remnant only.  The cut
$r_*>0.005$, shown as solid curves,
extends further into the outer halo.}
\label{fig:remnant2}
\end{figure}

It was found in Paper~1 that the addition of 1PN corrections to the
hydrodynamics has the effect of reducing mass shedding 
for mergers of NS with a $\Gamma=3$ EOS.  In Fig.~\ref{fig:remnant2},
we show the evolution of the remnant mass $M_r$ for runs C1 and D1, and find
the same behavior for a $\Gamma=2$ EOS.  We show two different density
cuts to highlight this behavior.  By taking all SPH particles with
a local density $r_*>0.005$, we extend our definition of the ``remnant''
far into the outer halo.
Instead, a density cut of $r_*>0.04$ includes
just the inner part of the remnant, and excludes any material that was
ejected into spiral arms.
In both runs mass shedding starts occuring approximately
at the time of maximum gravity
wave emission ($t=20$), and lasts for a total time $\delta t\simeq 10$, 
but in the PN
case the mass shedding rate is significantly smaller, leading to
approximately half the total amount of mass shedding as in 
the Newtonian case.

\begin{figure}
\centering \leavevmode \epsfxsize=7in \epsfbox{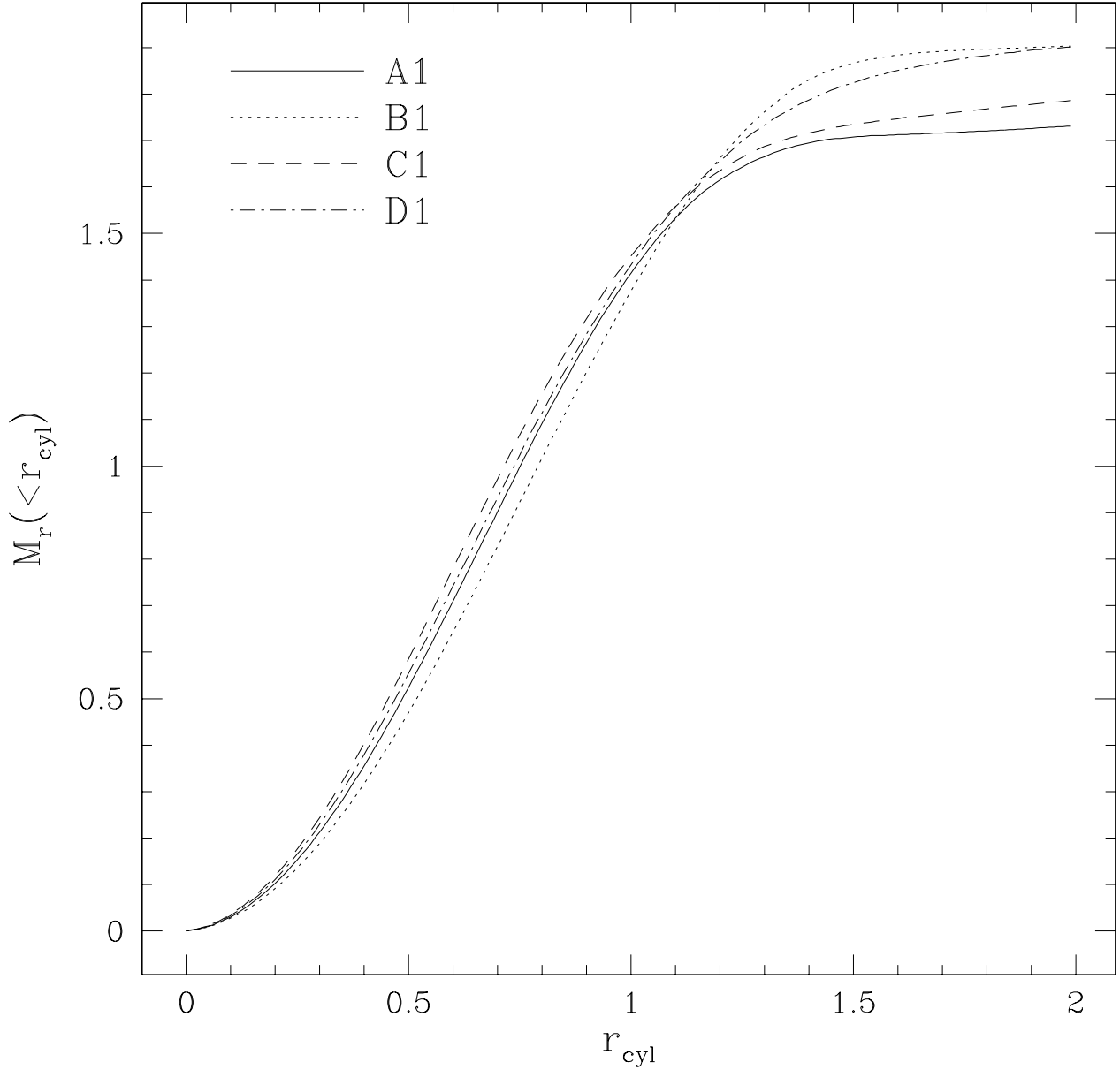}
\caption{Radial mass profiles of the final merger 
remnants in runs A1, B1, C1 and D1 (at $t=60$). Here $r_{\rm cyl}$ 
is the distance from the rotation axis. We
see that the internal structure is governed primarily by the EOS, with the
$\Gamma=2$ models slightly more centrally condensed than the $\Gamma=3$
models, although 1PN effects do decrease the enclosed mass at small
radii.  At larger radii, we find more mass contained in the PN
remnants, since much less mass has been ejected through
spiral arms.}
\label{fig:finalmass}
\end{figure}

In Fig.~\ref{fig:finalmass}, we show the radial mass profiles of the
inner remnants for the 
four synchronized $q=1$ runs.  
We see that the mass profile at small
radii is primarily determined by the EOS, with the $\Gamma=2$ models
showing slightly more central concentration, as would be expected,
although the inclusion of 1PN effects does decrease the enclosed mass
in a given cylinder.  At 
$r_{\rm cyl}\simeq 1.2$, however, we start to see significant differences 
between runs
with and without 1PN corrections.  In the Newtonian runs, matter is
ejected much more efficiently to large radii, and thus the final
mass of the inner remnant falls in the range $M_r=1.7-1.8$, whereas for the
PN runs it is $M_r\simeq 1.9$.  The remaining mass,
ejected through the spiral arms, forms the halo around the
inner remnant.  Here we can see from the slopes of the mass profiles
near $r_{\rm cyl}\simeq2$ that for a softer ($\Gamma=2$) EOS, the density in the
halo near the inner remnant is still significant, indicating
more mass in the inner region of the halo.
This is not a surprise since, as already noted by
RS2, a softer EOS produces 
wider spiral arms, which in turn dissipate more quickly and transport
material less efficiently out to very large radii.

In Table~\ref{table:runs},
we list in the last three columns
the total rest mass $M_r$, gravitating mass $M_{gr}$, and Kerr parameter 
$a_r$ of the inner remnant at $t=60$. Here the inner remnant
includes all SPH particles out to a radius $r=2$  (The results
are rather insensitive to the precise choice of density cut, since the
material in the halo is very tenuous).  As was done in Paper~1, the Kerr
parameter for all PN runs is taken to be $a_r=cJ_r/M_{gr}^2$, where
$J_r$ is the 1PN angular momentum of the remnant.

\begin{figure}
\centering \leavevmode \epsfxsize=7in \epsfbox{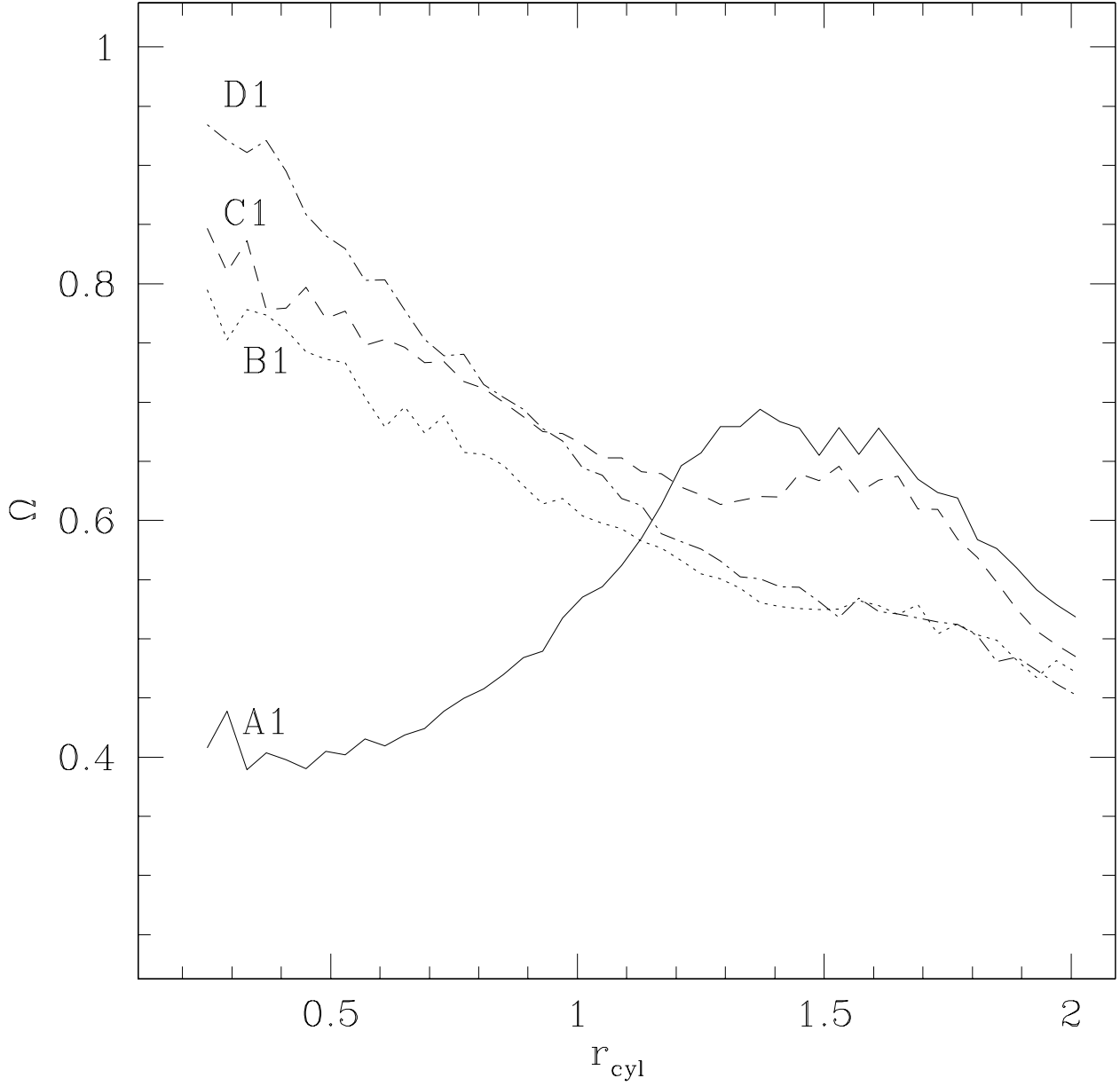}
\caption{Angular velocity profiles for the same merger remnants
shown in Fig.~22. We see that all are differentially rotating,
and all but the Newtonian $\Gamma=3$ remnant show a decrease in
angular velocity with increasing radius for $r_{\rm
cyl}\protect\lesssim 1.3$.  
In general, at larger radii,
the PN models show slower rotation, regardless of the EOS.}
\label{fig:finalvel}
\end{figure}

A comparison of the angular velocity profiles of the remnants 
is shown in Fig.~\ref{fig:finalvel}.  We see, quite
surprisingly, that the $\Gamma=3$ run without 1PN corrections (A1)
leads to
a different form than each of the other runs.  All are differentially
rotating, but in all other cases the angular velocity drops as a
function of distance away from the rotation axis, 
whereas in run A1 it increases monotonically out to $r_{\rm cyl}\simeq1.4$,
which is close to the surface of the inner remnant.  
Only run A1 agrees with previous purely Newtonian calculations (RS1, RS2), 
which found that the angular velocity increases with increasing distance from
the rotation axis. However, in agreement with all previous
studies, we find that the rotation profiles of all our merger remnants 
are pseudo-barotopic, i.e., $\Omega$
is a constant on cylinders.  Recent fully GR calculations have indicated
that differential rotation can increase very significantly the maximum 
stable mass of neutron stars \cite{DiffRot}, 
which makes it more likely that merger remnants can
be dynamically stable against collapse to a black hole.  However, 
differentially rotating
configurations could still be secularly unstable on a viscous timescale.

Run A1 is the only one that did not produce a second gravity wave luminosity
peak, suggesting that perhaps the combination of Newtonian
gravity and the stiffer EOS can lead to a lower central
angular momentum. In turn, this could suppress the quadrupole oscillations
shown in Fig.~\ref{fig:mom2}. We also note a general
difference in rotational velocity between the PN runs and
the Newtonian runs, with both N runs showing an increase in angular
velocity near the surface of the inner remnant, from $r_{\rm cyl}\simeq1.2-1.4$, 
whereas the PN runs show a steady decrease through this range and out
into the halo.  

\begin{figure}
\centering \leavevmode \epsfxsize=7in \epsfbox{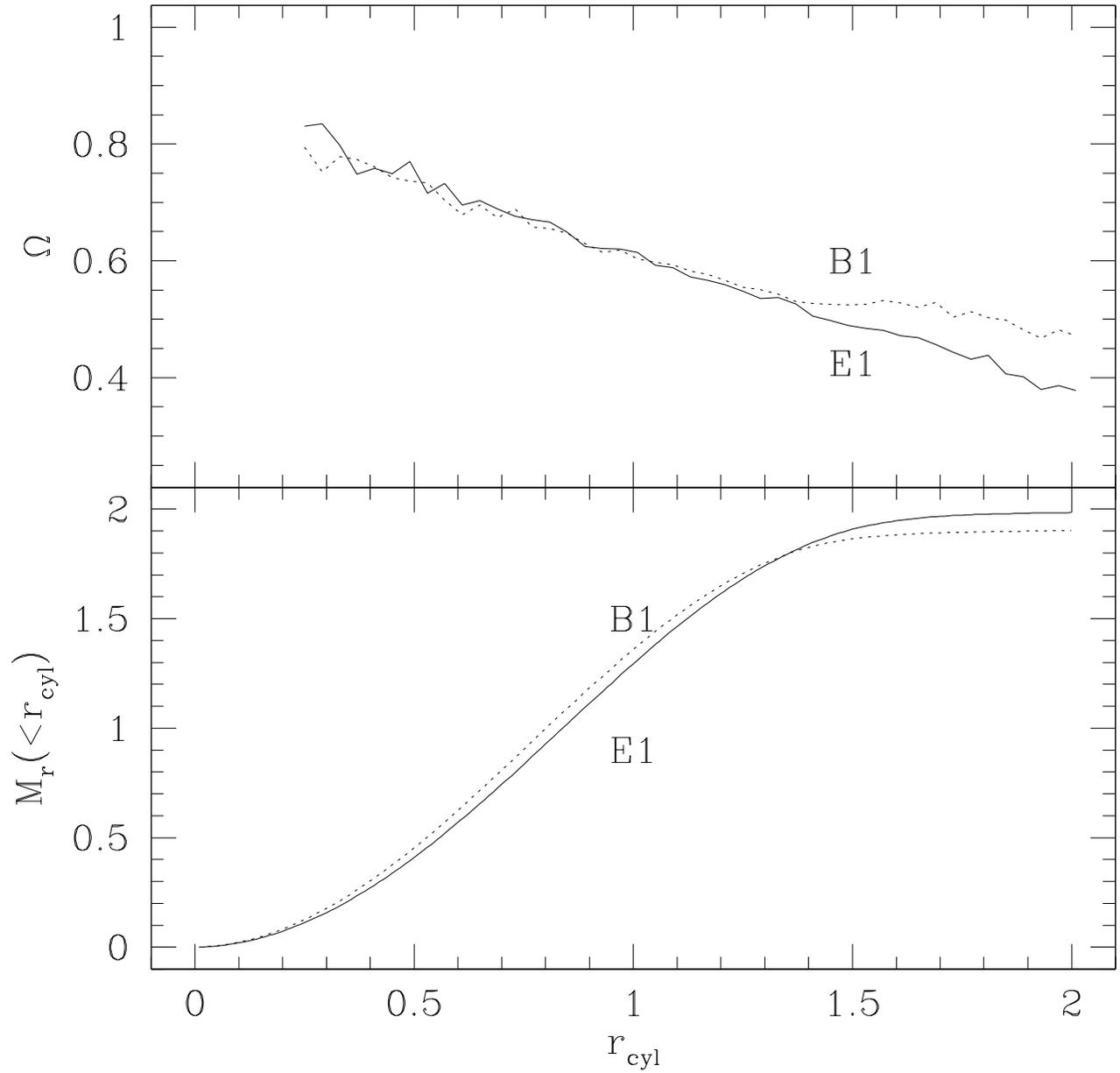}
\caption{Comparison of the merger remnants from runs B1 (synchronized
initial condition) and E1 (irrotational initial condition).  Both
runs have $\Gamma=3$ and 1PN corrections. Both the mass profiles and
the rotation profiles of the inner remnants appear to be rather 
independent of the initial NS spins.} 
\label{fig:finalirr}
\end{figure}

The final structure of the merger remnant appears remarkably independent
of the initial NS spins.
In Fig.~\ref{fig:finalirr} we show the mass and angular
velocity profiles of the merger remnant for the irrotational binary of run E1,
compared with that of the synchronized system in run B1.  We see that
the two merger remnants have very similar profiles. 
Both remnants are differentially
rotating with an angular velocity that decreases with increasing
distance from the axis. The
irrotational binary merger remnant has a slightly smaller angular velocity 
at large radii and a slightly less centrally concentrated mass profile.
Note also
that at large radii, the mass of the remnant for the irrotational
case is essentially $M_r=2$, i.e., very little
mass shedding occurred.

\section{Summary and directions for future work}

Using the Lagrangian SPH code described in Paper~1, which is complete to
1PN order and also includes all 2.5PN radiation reaction effects, we have
further investigated the properties of NS binary mergers in PN gravity.  
In addition to the method presented in Paper~1 for constructing PN
equilibrium initial conditions for synchronized
binaries, we have used the results of the analytic work by LRS based on
compressible ellipsoids to create realistic irrotational initial conditions.  
Our approach takes an exact, spherical PN equilibrium model
computed for an isolated NS, and deforms it according to the shape of
the irrotational Darwin-Riemann ellipsoid solution determined from the 
energy variational method of LRS. 

Using the hybrid formalism of Paper~1, in which radiation reaction is
treated realistically but 1PN effects are scaled down in amplitude to
remain numerically tractable, we have studied the effect of PN corrections
on several aspects of binary NS mergers.  We find that NS with a
softer EOS, modeled here as $\Gamma=2$ polytropes, produce higher
gravity wave luminosities during the merger than NS with stiffer EOS  
(modeled here and in Paper~1 as $\Gamma=3$ polytropes).  As was found 
in previous, purely Newtonian calculations, the final 
merger remnant produced when $\Gamma=2$ is an oblate spheroid, rather 
than a triaxial ellipsoid (obtained for $\Gamma=3$).  For calculations 
with and without 1PN effects, we find a
strong second peak of gravity wave luminosity for a $\Gamma=2$ EOS, in
contrast to the result obtained in 
Paper~1, where we found that for a $\Gamma=3$ EOS, a strong
second peak of emission is present only when 1PN effects are included.

By holding the EOS and the primary NS mass fixed while varying the 
secondary NS mass, we have studied the dependence of the gravity wave
emission on the binary mass ratio.  We find that the steep 
power-law scaling $h_{\rm max}\propto q^2$ for the maximum amplitude, derived
from purely Newtonian calculations in RS2 for a $\Gamma=2$ EOS, 
remains approximately
correct in PN gravity, for both $\Gamma=2$ and $\Gamma=3$ EOS.  
In both cases the dependence on $q$ appears slightly
steeper when 1PN effects are neglected (but while retaining 2.5PN effects).
The maximum gravity wave luminosity for PN mergers
follows a slightly steeper power law than the $L_{\rm max}\propto q^6$
found by RS2 for the $\Gamma=2$ EOS, but is significantly flatter for
the $\Gamma=3$ EOS.

Our coalescence calculation using an irrotational initial condition
shows clearly the development of a vortex sheet along the surface of contact.  
A turbulent region is seen until the time when the inner cores of the
two NS begin to merge, at which
point a stable, differentially rotating configuration is created.  While the
main gravity wave luminosity peak is smaller than for a synchronized
initial condition, the second peak is of similar relative amplitude and
coincident in time.  A strong third peak was not observed for the
irrotational case. The density and rotation profiles of the inner remnant
appear remarkably independent of the degree of synchronization of the
initial binary.

We find that, in general, the addition of 1PN effects decreases the
mass in the outer halo of the merger remnant, especially when an
irrotational initial condition is used.  For this latter case, almost
no mass at all is ejected through spiral arms during the merger.  
Since the 1PN corrections are artificially reduced in our calculations, 
we expect that for two NS with realistic parameters and an irrotational
initial configuration, no matter at all will be ejected.
When a synchronized
initial condition is used, we find that the halo around the remnant is
much denser and less extended for a softer NS EOS.

Our results have potentially important consequences for theoretical
models of GRBs based on NS binary mergers \cite{RJTS,GRB1}.  
Although current models
for {\it long\/} bursts and their associated observed afterglows
favor an origin in massive star collapse, NS mergers remain an
attractive source at least for the separate class of {\it short\/} bursts 
\cite{GRB2}.
Currently the most popular models all assume that a binary NS merger leads
to the formation of a rapidly rotating Kerr BH surrounded by a torus
of ejected material. 
Energy can then be extracted either from the rotation of the BH or from
the material in the torus so that, with sufficient beaming, the
gamma-ray fluxes observed from even the most distant GRBs can be
explained \cite{GRB3}. 
However, our results suggest that the 
merger of two $1.4\,M_\odot$ NS with stiff EOS forms an object that will 
{\it not\/} 
collapse to a BH on a dynamical time (see Paper I, Sec.\ IIID ), in agreement
with the preliminary full GR calculations of Shibata and Uryu \cite{Shi2}.
Moreover, for irrotational binaries (containing two slowly spinning NS; see
Sec.\ \ref{sec:irrotres}), we predict that {\it no matter\/} will be ejected
during the merger, implying that no outer torus will form around the 
central core (see Sec.\ \ref{sec:remnant}). 
Even if the central core were to collapse to form
a Kerr BH (which is not prevented by rotation, since $a_r < 1$; see Table 
\ref{table:runs}),
energy could not be extracted to power a GRB.

Our future work using PN SPH will focus on the more realistic
irrotational initial conditions introduced in this paper, while
again varying the other parameters of the problem such as the binary
mass ratio and the NS EOS. We will also study the dependence of our
numerical results on the spatial resolution of the calculations by
systematically varying the number of SPH particles from $N\sim10^4$
all the way to $N\sim10^6$ (which will take full advantage of our
highly optimized parallel SPH code). We will try to establish that our
conclusions about the gravity wave signal and final merger remnant
configuration are free from numerical artifacts, in spite of the
difficulties described in Sec.~\ref{sec:irrotres}.  Increased resolution
should also yield a better understanding of the nature of the vortex
sheet formed upon first contact of the binary components, and of the
details of the transition to a differentially rotating configuration.
We also plan to study the properties of the gravitational wave signals
in Fourier space (as was done for the first time by Zhuge {\it et al.} 
\cite{Zhu1,Zhu2} based on their Newtonian SPH calculations), since the
spectral information is crucial to the interpretation of future results
from LIGO and similar instruments.  We will study systematically the 
dependence of the power spectrum of the gravitational radiation on 
the binary mass ratio, and on the NS EOS, as well as
investigate the effects of 1PN corrections on the power spectrum.

\acknowledgements
This work was supported in part by NSF Grants AST-9618116 and PHY-0070918 
and NASA ATP Grant NAG5-8460. 
F.A.R.\ was supported in part by an Alfred P.\ Sloan Research Fellowship.
The computations were 
supported by the National Computational Science Alliance under grant
AST980014N and utilized the NCSA SGI/CRAY Origin2000.

\newpage

\newpage

\newpage

\newpage

\newpage

\newpage

\newpage

\newpage

\newpage

\newpage

\newpage

\newpage

\newpage

\newpage

\newpage

\newpage

\newpage

\newpage

\newpage

\newpage

\newpage

\newpage

\newpage

\newpage
\begin{table}
\caption{Properties of the $\Gamma=2$ polytropic NS sequence. Units are
defined such that $G=M=R=1$.  The polytropic constant $k$ is defined by
$P=kr_*^{\Gamma}$, and $k_N$ is the Newtonian value, which is used in the
Lane-Emden equation.  The central density $(r_*)_c$ is in units of $M/R^3$, 
while the central pressure ratio $(P/r_*c^2)_c$ and central
gravitational potential ratio $(U_*/c^2)_c$ are dimensionless.}
\label{table:single}
\begin{tabular}{ccccc}
$1/c^2$&$k/k_N$&$(r_{*})_c$&$(P/r_*c^2)_c$&$(U_*/c^2)_c$\\
\tableline
 & & $\Gamma=2$ & &\\
\tableline
0.01 & 1.202 & 0.6549 & 0.0050 & 0.0187 \\
0.02 & 1.260 & 0.6670 & 0.0107 & 0.0376 \\
0.03 & 1.308 & 0.6963 & 0.0174 & 0.0571 \\
0.04 & 1.375 & 0.7173 & 0.0251 & 0.0767 \\
0.05 & 1.439 & 0.7490 & 0.0343 & 0.0968  \\
\end{tabular}
\end{table}

\begin{table}
\caption{Properties of the PN polytropic models for NS of
varying masses.  Here the
equation of state remains fixed for a given adiabatic index.  All models
have $1/c^2=0.05$.  Quantities are defined as in Table~\ref{table:single}.}
\label{table:single2}
\begin{tabular}{ccccc}
M & R &$(r_{*})_c$&$(P/r_*c^2)_c$&$(U_*/c^2)_c$\\
\tableline
 & & $\Gamma=2, k/k_N=1.439$ & &\\
\tableline
1.0 & 1.0 & 0.7490 & 0.0343 & 0.0968 \\
0.8 & 1.025 & 0.5311 & 0.0243 & 0.0747 \\
\tableline
 & & $\Gamma=3, k/k_N=1.918$ & &\\
\tableline
1.0 & 1.0 & 0.4051 & 0.0397 & 0.0813 \\
0.9 & 0.986 & 0.3781 & 0.0349 & 0.0757 \\
0.8 & 0.969 & 0.3501 & 0.0299 & 0.0683 \\
\end{tabular}
\end{table}

\newpage
\begin{table}
\caption{Selected quantities from each of the simulations.  Note that
only runs A1, B1, C1, D1, D2 and E1 were extended until a stable 
final remnant was
formed.  All other runs were terminated shortly after the first gravity
wave luminosity peak.  Here $t_0$ is the time at which the run was started (see
Sec.~\protect\ref{sec:sumcalc}), and $\theta_{\rm lag}$ is the lag angle at first
contact, given for
both NS (see Sec.~\protect\ref{sec:results}).  Quantities involving the first
and second gravity wave luminosity peaks are labeled with superscripts
$(1)$ and $(2)$.
All SPH particles with $r<2.0$ are used to calculate the final rest mass $M_r$,
gravitational mass $M_{gr}$, and Kerr parameter $a_r$ of the final remnant 
(at $t=60$; see Sec.~\protect\ref{sec:remnant}).}
\label{table:runs}
\begin{tabular}{ccc|l|cc|ccc|ccc}
Run & q & $t_0$ & $\theta_{lag} (deg)$ & $c^5L^{(1)}_{GW}$ & $c^4(dh^{(1)}_{\rm max})$ & 
$c^5L^{(2)}_{GW}$ & $c^4(dh^{(2)}_{\rm max})$ & $t^{(2)}_{\rm max}$ &
$M_r$ & $M_{gr}$ & $a_r$ \\ 
\tableline
\multicolumn{11}{c}{N, $\Gamma=3.0$, Synchronized, $r_0=3.1$} \\
\tableline
A1 & 1.00 & -13 & 4.2 & 0.4061 & 2.2469 & 0.0615 & 0.6421 & 33 & 1.730
& N/A & 0.469 \\
A2 & 0.95 & -15 & 3.0, 3.5 & 0.3284 & 2.0505 & & & & & & \\ 
A3 & 0.90 & -18 & 3.0, 3.7 & 0.2617 & 1.8439 & & & & & & \\
A4 & 0.85 & -22 & 3.0, 4.8 & 0.2045 & 1.6545 & & & & & & \\
A5 & 0.80 & -27 & 3.0, 5.0 & 0.1567 & 1.4850 & & & & & & \\
\tableline
\multicolumn{11}{c}{PN, $\Gamma=3.0$, Synchronized, $r_0=4.0$} \\
\tableline
B1 & 1.00 & 0   & 11.2 & 0.3317 & 2.0549 & 0.2253 & 1.2129 & 33 &
1.902 & 1.847 & 0.716 \\
B2 & 0.90 & -2  & 10.3, 11.1 & 0.2280 & 1.7999 & & & & & & \\
B3 & 0.80 & -4  & 9.1, 12.1 & 0.1216 & 1.5291 & & & & & & \\
\tableline
\multicolumn{11}{c}{N, $\Gamma=2.0$, Synchronized, $r_0=2.9$}\\
\tableline
C1 & 1.00 & -13 & 4.0 & 0.5837 & 2.3830 & 0.0721 & 0.5668 & 38 & 1.785
& N/A & 0.590 \\
C2 & 0.80 & -13 & 5.1, 11.5 & 0.0891 & 1.3635 & & & & & & \\
\tableline
\multicolumn{11}{c}{PN, $\Gamma=2.0$, Synchronized, $r_0=4.0$}\\
\tableline
D1 & 1.00 & -2  & 11.0 & 0.4510 & 2.0890 & 0.1442 & 0.8258 & 29 &
1.900 & 1.747 & 0.832 \\
D2 & 0.80 & 2   & 5.2, 10.1 & 0.0885 & 1.4456 & 0.0530 & 1.0896 & 31 &
1.646 & 1.529 &  0.822 \\
\tableline
\multicolumn{11}{c}{PN, $\Gamma=3.0$, Irrotational, $r_0=4.0$}\\
\tableline
E1 & 1.00 & 3   & 14.5 & 0.2419 & 1.9093 & 0.1475 & 1.0200 & 33 &
1.983 & 1.913 & 0.765 \\
\end{tabular}
\end{table}


\begin{thebibliography}{DUM}
\bibitem{1} A.\ Abramovici {\it et al.}, Science {\bf 256}, 325
(1992); A.\ Abramovici {\it et al.}, Phys. Lett. A {\bf 218}, 157 (1996).

\bibitem{2} C.\ Bradaschia {\it et al.}, Nucl. Instrum. Methods {\bf
 A289}, 518 (1990);  B.\ Caron {\it et al.}, Class. Quantum Grav. {\bf
14}, 1461 (1997).

\bibitem{3} J.\ Hough, in {\it Proceedings of the Sixth Marcel Grossmann
Meeting}, edited by H.~Sato and T.~Nakamura (World Scientific,
Singapore, 1992), p. 192.

\bibitem{3a} K.\ Danzmann, in {\it Relativistic Astrophysics},
proceedings of the 162nd W.E. Heraeus Seminar, edited by H.\ Riffert
{\it et al.} (Wiesbaden: Vieweg Verlag, 1998), p.48.

\bibitem{4} K.\ Kuroda {\it et al.}, in {\it Proceedings of the
International Conference on Gravitational Waves: Sources and Detectors}, 
edited by I.~Ciufolini and F.~Fidecard (World Scientific, 1997),
p.100.

\bibitem{15} B.J.\ Meers, Phys. Rev. D {\bf 38}, 2317 (1988); K.A.\ Strain and
B.J.\ Meers, Phys. Rev. Lett. {\bf 66}, 1391 (1991).

\bibitem{Nak1} K.\ Oohara and T.\ Nakamura, Prog. Theor. Phys. {\bf
82}, 535 (1989); T.\ Nakamura and K.\ Oohara, {\it ibid.} {\bf 82},
1066 (1989); K.\ Oohara and T.\ Nakamura, {\it ibid.} {\bf
83}, 906 (1990); T.\ Nakamura and K.\ Oohara, {\it ibid.} {\bf 86},
73 (1991). 

\bibitem{Nak2}  M.\ Shibata, K.\ Oohara, and T.\ Nakamura,
Prog. Theor. Phys. {\bf 88}, 1079 (1992); {\bf 89}, 809 (1993).

\bibitem{RS13} F.A.\ Rasio and S.L.\ Shapiro, Astrophys. J. {\bf 401},
226 (1992) [RS1]; {\bf 432}, 242 (1994) [RS2]; {\bf 438}, 887 (1995)
[RS3].

\bibitem{Zhu1} X.\ Zhuge, J.\ Centrella, and S.\ McMillan, Phys. Rev. D
{\bf 50}, 6247 (1994).

\bibitem{Zhu2} X.\ Zhuge, J.\ Centrella, and S.\ McMillan, Phys. Rev. D
{\bf 54}, 7261 (1996).

\bibitem{Dav} M.B.\ Davies, W.\ Benz, T.\ Piran, and F.K.\ Thielemann,
Astrophys. J. {\bf 431}, 742 (1994)

\bibitem{Ros1} S.\ Rosswog {\it et
al.}, Astron. Astrophys. {\bf 341}, 499 (1999).

\bibitem{Ros2} S.\ Rosswog, M.B.\ Davies, F.-K.\ Thielemann, and T.\
Piran, Astron. Astrophys. {\bf 360}, 171 (2000).

\bibitem{New} K.C.B.\ New and J.E.\ Tohline, Astrophys. J. {\bf 490}, 311
(1997). 

\bibitem{Swe} F.D.\ Swesty, E.Y.M.\ Wang, and A.C.\ Calder,
Astrophys. J. (to be published), astro-ph/9911192.

\bibitem{RJS} M.\ Ruffert, H.-Th.\ Janka, and
G.\ Sch\"{a}fer, Astron. Astrophys. {\bf 311}, 532 (1996).

\bibitem{RJTS} M.\ Ruffert,
H.-Th.\ Janka, K.\ Takahashi, and G.\ Sch\"{a}fer, Astron. Astrophys.
{\bf 319}, 122 (1997). 

\bibitem{RRJ} M.\ Ruffert, M.\ Rampp, and H.-Th.\ Janka,
Astron. Astrophys. {\bf 321}, 991 (1997).

\bibitem{Nak3} K.\ Oohara and T.\ Nakamura, Prog. Theor. Phys. {\bf
88}, 307 (1992).

\bibitem{FR1} J.A.\ Faber and F.A.\ Rasio, Phys. Rev. D, {\bf 62}, 064012 
(2000) [Paper 1]. 

\bibitem{Ayal} S.\ Ayal {\it et al.}, Astrophys. J. (submitted),
astro-ph/9910154. 

\bibitem{BDS} L.\ Blanchet, T.\ Damour, and G.\ Sch\"{a}fer, Mon.
Not. R. Astron. Soc. {\bf 242}, 289 (1990) [BDS].

\bibitem{16} T.W.\ Baumgarte, S.A.\ Hughes, and S.L.\ Shapiro,
Phys. Rev. D {\bf 60}, 087501 (1999); W.\ Landry and S.A.\ Teukolsky,
Phys. Rev. D (to be published), gr-qc/9912004; 
M.\ Miller, W.\ Suen, and M.\ Tobias, Phys. Rev. D (to be
published), gr-qc/9910022; E.\ Seidel, in {\it Relativistic Astrophysics},
proceedings of the 162nd W.E. Heraeus Seminar, edited by H.\ Riffert
{\it et al.} (Wiesbaden: Vieweg Verlag, 1998), p.229.

\bibitem{Shi1} M.\ Shibata, Phys. Rev. D {\bf 60}, 104052 (1999).

\bibitem{Shi2} M.\ Shibata and K.\ Uryu, Phys. Rev. D. {\bf 61}, 
064001 (2000). 

\bibitem{Tho} S.E.\ Thorsett and D. Chakrabarty, D. Astrophys. J. {\bf
512}, 288 (1999).

\bibitem{Tay} J.H.\ Taylor and J.M.\ Weisberg, Astrophys. J. {\bf 345},
434 (1989).

\bibitem{21} W.T.S.\ Deich and S.R.\ Kulkarni, in {\it IAU Symp. 165:
Compact Stars in Binaries}, edited by J.\ van Paradijs, E.P.J.\ van
den Heuvel, and E.\ Kuulkers (Kluwer Academic Publishers, Dordrecht, 1995).

\bibitem{Wol} A. Wolszczan, Nature {\bf 350}, 688 (1991).

\bibitem{NSmass}  F.X.\ Timmes, S.E.\ Woosley, and T.A.\ Weaver,
Astrophys. J. {\bf 457}, 834 (1996); V.\ Kalogera, and C.L.\ Fryer,
Astrophys. J. (submitted), astro-ph/9911312.

\bibitem{JERF} H.-Th.\ Janka, Th.\ Eberl, M.\ Ruffert, and C.L.\ Fryer,
Astrophys. J. Lett. {\bf 527}, L39 (1999).

\bibitem{LK} W.H.\ Lee and W.L.\ Kluzniak, Astrophys. J. {\bf 526},
178 (1999); Mon. Not. R. Astron. Soc. {\bf 308}, 780 (1999).

\bibitem{Irrot} C.S. Kochanek, Astrophys. J. {\bf 398}, 234 (1992);
L. Bildsten and C. Cutler, {\it ibid.} {\bf 400}, 175 (1992).

\bibitem{LRS1} D. Lai, F.A.\ Rasio, and S.L. Shapiro, 
Astrophys. J. Lett. {\bf 406}, L63 (1993).

\bibitem{LRS3} D. Lai, F.A.\ Rasio, and S.L. Shapiro, 
Astrophys. J. {\bf 420}, 811 (1994).

\bibitem{Note1} Note that BDS use
$r$ to denote the mass-energy density, while we use $r$ in this
paper (and in Paper~1) to denote the radial distance from the
origin of our coordinate system. The two BDS quantities $r$ and $r_*$
become equal in the Newtonian limit, but differ at the 1PN
level. Similarly, $M$, the total baryonic mass defined in
terms of $r_*$, will
differ by 1PN terms from $M_g$, the total gravitating mass.
See Eqs.~3 and 4 of Paper~1 for a more detailed discussion.

\bibitem{ST} S.L.\ Shapiro and S.\ Teukolsky, {\it Black Holes, White
Dwarfs, and Neutron Stars} (Wiley, New York, 1983).

\bibitem{Lat} J.M.\ Lattimer and D.\ Swesty, Nucl. Phys. {\bf A535}, 331
(1991). 

\bibitem{Bay} G. Baym, in {\it Neutron Stars: Theory and Observation},
edited by J.\ Ventura and D.\ Pines (Dordrecht, Kluwer, 1991), p. 21.

\bibitem{Akm} A.\ Akmal, V.R.\ Pandharipande, and D.G.\ Ravenhall,
Phys. Rev. C {\bf 58}, 1804 (1998). 

\bibitem{Kaon} N.K.\ Glendenning, Phys. Rev. D {\bf 46}, 1274 (1992);
G.E.\ Brown and H.A.\ Bethe, Astrophys. J. {\bf 423}, 659 (1994).

\bibitem{LRS} J.C.\ Lombardi, F.A.\ Rasio, and S.L.\ Shapiro,
Phys. Rev. D {\bf 56}, 3416 (1997) [LRS].

\bibitem{Chandra} S.\ Chandrasekhar, {\it Ellipsoidal Figures of
Equilibrium; Revised Dover Edition} (Yale University Press, New Haven,
1987).

\bibitem{Note2} Note that, recently, many other groups have obtained
approximate solutions for irrotational close binary configurations
containing polytropes, both in Newtonian and PN gravity, and in full GR. 
See K.\ Taniguchi and T.\ Nakamura, Phys. Rev. D (to be published),
astro-ph/0004010; Phys. Rev. Lett. {\bf 84}, 581 (2000);
K.\ Uryu and Y.\ Eriguchi, Phys. Rev. D {\bf 61}, 124023 (2000);
S.\ Bonazzola, E.\ Gourgoulhon, and J.-A.\ Marck,
Phys. Rev. Lett. {\bf 82} 892, (1999);
K.\ Taniguchi, Prog. Theor. Phys. {\bf 101}, 283 (1999);
S.A.\ Teukolsky, Astrophys. J. {\bf 504}, 402 (1998).

\bibitem{RS94} F.A.\ Rasio and S.L.\ Shapiro, in {\it IAU Symp. 165:
Compact Stars in Binaries}, edited by J.\ van Paradijs, E.P.J.\ van
den Heuvel, and E.\ Kuulkers (Kluwer Academic Publishers, Dordrecht,
1995);  F.A.\ Rasio and S.L.\ Shapiro, Class. Quantum Grav. {\bf 16},
1 (1999).

\bibitem{DiffRot} T.W.\ Baumgarte, S.L.\ Shapiro, and M. Shibata,
Astrophys. J. Lett. {\bf 528}, L29 (2000).

\bibitem{GRB1} D.\ Eichler, M.\ Livio, T.\ Piran, and D.N.\ Schramm, Nature 
{\bf 340}, 126 (1989); R.\ Narayan, B.\ Paczy\'nski, and T.\ Piran, 
Astrophys. J. Lett. {\bf 395}, L83 (1992); P.\ Meszaros and M.J.\ Rees, 
Astrophys. J. {\bf 397}, 570 (1992); V.M.\ Lipunov {\it et al.}, {\it ibid.} 
{\bf 454}, 593 (1995); J.S.\ Bloom, S.\ Sigurdsson, and O.R.\ Pols, Mon. Not. 
R. Astron. Soc. {305}, 763 (1999).

\bibitem{GRB2} C.\ Kouveliotou {\it et al.}, Astrophys. J. Lett. {\bf 413}, 
L101 (1993); A.I.\ MacFayden and S.E.\ Woosley, Astrophys. J. {\bf 524}, 
262 (1999); P.\ Meszaros, Nucl. Phys. Proc. Suppl. {\bf 80}, 63 (2000);
M.\ Livio and E.\ Waxman, Astrophys. J. {\bf 538}, 187 (2000).

\bibitem{GRB3} P.\ Meszaros, M.J. Rees, and R.A.M.J.\ Wijers, New Astron. 
{\bf 4}, 303 (1999).

\end{thebibliography}
\end{document}